\documentstyle[aps,twocolumn]{revtex}

\input{epsf}

\title{Secure quantum key distribution using squeezed states\thanks{CALT-68-2274}}
\author{Daniel Gottesman,$^{(1)}$\thanks{\tt gottesma@microsoft.com} and John Preskill$^{(2)}$\thanks{\tt
preskill@theory.caltech.edu}}

\address{$^{(1)}$Microsoft Corporation, One Microsoft Way, Redmond, WA 98052, USA\\ 
$^{(2)}$California Institute of Technology, 
Pasadena, CA 91125, USA }

\begin{document}

\maketitle

\begin{abstract}
We prove the security of a quantum key distribution scheme based on transmission of squeezed quantum states of a harmonic oscillator. Our proof employs quantum error-correcting codes that encode a finite-dimensional quantum system in the infinite-dimensional Hilbert space of an oscillator, and protect against errors that shift the canonical variables $p$ and $q$. If the noise in the quantum channel is weak, squeezing signal states by 2.51 dB (a squeeze factor $e^r=1.34$) is sufficient in principle to ensure the security of a protocol that is suitably enhanced by classical error correction and privacy amplification. Secure key distribution can be achieved over distances comparable to the attenuation length of the quantum channel.
\end{abstract}

\section{Introduction}
Two of the most important ideas to emerge from recent studies of quantum information are the concepts of quantum error correction and quantum key distribution.  Quantum error correction allows us to protect unknown quantum states from the ravages of the environment.  Quantum key distribution allows us to conceal our private discourse from potential eavesdroppers. 

In fact these two concepts are more closely related than is commonly appreciated.  A quantum error correction protocol must be able to reverse the effects of both bit flip errors, which reflect the polarization state of a qubit about the $x$-axis, and phase errors, which reflect the polarization about the $z$-axis.  By reversing both types of errors, the protocol removes any entanglement between the protected state and the environment, thus restoring the purity of the state. 

In a quantum key distribution protocol, two communicating parties verify that qubits polarized along both the $x$-axis and the $z$-axis can be transmitted with an acceptably small probability of error.  An eavesdropper who monitors the $x$-polarized qubits would necessarily disturb the $z$-polarized qubits, while an eavesdropper who monitors the $z$-polarized qubits would necessarily disturb the $x$-polarized qubits.  Therefore, a successful verification test can show that the communication is reasonably private, and the privacy can then be amplified via classical protocols.

In quantum key distribution, the eavesdropper collects information by entangling her probe with the transmitted qubits. Thus both error correction and key distribution share the goal of protecting quantum states against entanglement with the outside world.

Recently, this analogy between quantum error correction and quantum key distribution has been sharpened into a precise connection, and used as the basis of a new proof of security against all possible eavesdropping strategies \cite{shor_pres}.  Earlier proofs of security (first by Mayers \cite{mayers}, and later by Biham {\it et al.} \cite{biham}) made no explicit reference to quantum error correction; nevertheless, the connection between quantum error correction and quantum key distribution is a powerful tool, enabling us to invoke the sophisticated formalism of quantum error-correcting codes in an analysis of the security of quantum key distribution protocols.  

Also recently, new quantum error-correcting codes have been proposed that encode a finite-dimensional quantum system in the infinite-dimensional Hilbert space of a quantum system described by continuous variables \cite{gott_kit_pres}. In this paper, we will apply these new codes to the analysis of the security of quantum key distribution protocols. By this method, we prove the security of a protocol that is based on the transmission of squeezed quantum states of an oscillator. The protocol is secure against all eavesdropping strategies allowed by the principles of quantum mechanics.

In our protocol, the sending party, Alice, chooses at random to send either a state with a well defined position $q$ or momentum $p$. Then Alice chooses a value of $q$ or $p$ by sampling a probability distribution, prepares a narrow wave packet centered at that value, and sends the wave packet to the receiving party, Bob. Bob decides at random to measure either $q$ or $p$. 
Through public discussion, Alice and Bob discard their data for the cases in which Bob measured in a different basis than Alice used for her preparation, and retain the rest. To correct for possible errors, which could be due to eavesdropping, to noise in the channel, or to intrinsic imperfections in Alice's preparation and Bob's measurement, Alice and Bob apply a classical error correction and privacy amplification scheme, extracting from the raw data for $n$ oscillators a number $k<n$ of key bits. 

Alice and Bob also sacrifice some of their data to perform a verification test to detect potential eavesdroppers. When verification succeeds, the probability is exponentially small in $n$ that any eavesdropper has more than an exponentially small amount of information about the key. Intuitively, this protocol is secure because an eavesdropper who monitors the observable $q$ necessarily causes a detectable disturbance of the complementary observable $p$ (and vice versa). 

Since preparing squeezed states is technically challenging, it is important to know how much squeezing is needed to ensure the security of the protocol. The answer depends on how heavily the wave packets are damaged during transmission. When the noise in the channel is weak, we show that it suffices in principle for the squeezed state to have a width smaller by the factor $e^{-r}=.749$ than the natural width of a coherent state (corresponding to an improvement by 2.51 dB in the noise power for the squeezed observable, relative to vacuum noise). It is also important to know that security can be maintained under realistic assumptions about the noise and loss in the channel. Our proof of security applies if the protocol is imperfectly implemented, and shows that secure key distribution can be achieved over distances comparable to the attenuation length of the channel. Squeezed-state key distribution protocols may have some practical advantages over single-qubit protocols, in that neither single-photon sources nor very efficient photodetectors are needed.

Key distribution protocols using continuous variable quantum systems have been described previously by others \cite{ralph,hillery,reid}, but ours is the first complete discussion of error correction and privacy amplification, and the first proof of security against arbitrary attacks.

In \S\ref{sec:codes} we review continuous variable quantum error-correcting codes  \cite{gott_kit_pres} and in \S\ref{sec:qkd_qecc} we review the argument \cite{shor_pres} exploiting quantum error-correcting codes to demonstrate the security of the BB84 quantum key distribution scheme \cite{bb84}. This argument is extended to apply to continuous variable key distribution schemes in \S\ref{sec:qkd_cont} and \S\ref{sec:secure}. Estimates of how much squeezing is required to ensure security of the protocol are presented in \S\ref{sec:gaussian}. The effects on security of losses due to photon absorption are analyzed in \S\ref{sec:losses}, and \S\ref{sec:conclude} contains conclusions.

\section{Codes for continuous quantum variables}
\label{sec:codes}

We begin by describing codes for continuous quantum variables \cite{gott_kit_pres}. The two-dimensional Hilbert space of an encoded qubit embedded in the infinite-dimensional Hilbert space of a system described by canonical variables $q$ and $p$ can be characterized as the simultaneous eigenspace of the two commuting operators
\begin{equation}
\label{stabilizer}
S_q=e^{i(2\sqrt{\pi})q}~,\quad S_p=e^{-i(2\sqrt{\pi})p}~,
\end{equation}
the code's ``stabilizer generators.'' If the eigenvalues are $S_q=S_p=1$, then the allowed values of $q$ and $p$ in the code space are integer multiples of $\sqrt{\pi}$, and the codewords are invariant under shifts in $q$ or $p$ by integer multiples of $2\sqrt{\pi}$. 
Thus an orthogonal basis for the encoded qubit can be chosen as
\begin{eqnarray}
\label{codewords}
|\bar 0\rangle&\propto & \sum_{s=-\infty}^{\infty} |q= (2s)\cdot \sqrt{\pi}\rangle\nonumber\\
&\propto & \sum_{s=-\infty}^{\infty} |p= s\cdot \sqrt{\pi}\rangle~,\nonumber\\
|\bar 1\rangle&\propto &\sum_{s=-\infty}^{\infty} |q= (2s+1)\cdot \sqrt{\pi}\rangle\nonumber\\
&\propto& \sum_{s=-\infty}^{\infty} (-1)^s|p=s\cdot \sqrt{\pi}\rangle~.
\end{eqnarray}
The operators
\begin{equation}
\bar Z\equiv  e^{i(\sqrt{\pi})q}~, \quad \bar X \equiv \quad e^{-i(\sqrt{\pi})p}~,
\end{equation}
commute with the stabilizer generators and so preserve the code subspace; they act on the basis eq.~(\ref{codewords}) according to 
\begin{eqnarray}
\label{encoded_xz}
\bar Z: & |\bar 0\rangle \to |\bar 0\rangle~, \quad & |\bar 1\rangle \to - |\bar 1\rangle~,\nonumber\\
\bar X: & |\bar 0\rangle \to |\bar 1\rangle~, \quad & |\bar 1\rangle \to  |\bar 0\rangle~.
\end{eqnarray}

This code is designed to protect against errors that induce shifts in the values of $q$ and $p$. To correct such errors, we measure the values of the stabilizer generators to determine the values of $q$ and $p$ modulo $\sqrt{\pi}$, and then apply a shift transformation to adjust $q$ and $p$ to the nearest integer multiples of $\sqrt{\pi}$. If the errors induce shifts $\Delta q$, $\Delta p$ that satisfy
\begin{equation}
|\Delta q| < \sqrt{\pi}/2~,\quad |\Delta p| < \sqrt{\pi}/2~,
\end{equation}
then the encoded state can be perfectly restored.

A code that protects against shifts is obtained for any choice of the eigenvalues of the stabilizer generators. The code with
\begin{equation}
S_q=e^{2\pi i\phi_q}~,\quad S_p=e^{-2\pi i \phi_p}~,
\end{equation}
can be obtained from the $\phi_q=\phi_p=0$ code by applying the phase space translation operator
\begin{equation}
e^{i\sqrt{\pi}(q\phi_p)}e^{-i\sqrt{\pi}(p\phi_q)}~;
\end{equation} 
the angular variables $\phi_{q}$ and $\phi_p \in (-1/2,1/2]$ denote the allowed values of $q/\sqrt{\pi}$ and $p/\sqrt{\pi}$ modulo an integer. In this code space, the encoded operations $\bar Z$ and $\bar X$ (which square to the identity) can be chosen to be
\begin{equation}
\bar Z(\phi_q)= e^{i\sqrt{\pi}(q-\phi_q\sqrt{\pi})}~,\quad \bar X(\phi_p)= e^{-i\sqrt{\pi}(p-\phi_p\sqrt{\pi})} ~.
\end{equation}

The code with stabilizer eq.~(\ref{stabilizer}) can be generalized in a variety of ways \cite{gott_kit_pres}. For example, we can increase the dimension of the protected code space, and we can modify the code to protect against shifts that are asymmetric in $q$ and in $p$. If we choose the stabilizer to be 
\begin{eqnarray}
\label{n_and_alpha}
S_q(n,\alpha)&=&\exp\left[{i(\sqrt{2\pi d})\cdot(q/\alpha)}\right]~,\nonumber\\
S_p(n,\alpha)&=&\exp\left[{-i(\sqrt{2\pi d})\cdot (p\alpha)}\right]~,
\end{eqnarray}
where $d$ is a positive integer and $\alpha$ is a positive real number, then the code has dimension $d$ and protects against shifts that satisfy
\begin{equation}
|\Delta q| < {\alpha\over 2}\cdot \sqrt{2\pi\over d}~,\quad |\Delta p| < {1\over 2\alpha}\cdot\sqrt{2\pi\over d}~.
\end{equation}

The codewords eq.~(\ref{codewords}) are nonnormalizable states, infinitely ``squeezed'' in $q$ and $p$. In practice, we must always work with normalizable finitely squeezed states. For example, a Gaussian approximation $|\tilde 0\rangle$ to the ideal codeword $|\bar 0\rangle$ of the $d=2$, $\alpha=1$ code, characterized by squeezing parameters $\Delta_q,\Delta_p << 1$,  is 
\begin{eqnarray}
|\tilde 0\rangle
 &\approx& \left({4\over \pi}\right)^{1/4}\int_{-\infty}^{\infty} dq \, |q\rangle \, e^{-{1\over 2}\left(\Delta_p^2\right) q^2} \nonumber\\
& \times &\sum_{s=-\infty}^{\infty}e^{-{1\over 2}(q - 2s\sqrt{\pi})^2/\Delta_q^2}\nonumber\\
 &\approx & {1\over \pi^{1/4}}\int_{-\infty}^{\infty} dp \, |p\rangle \, e^{-{1\over 2}\left(\Delta_q^2\right)p^2} \nonumber\\
&\times& \sum_{s=-\infty}^{\infty}e^{-{1\over 2}(p - s\sqrt{\pi})^2/\Delta_p^2}~;
\end{eqnarray}
the approximate codeword $|\tilde 0\rangle$ can be obtained by subjecting $|\bar 0\rangle$ to shifts in $q$ and $p$ governed by  Gaussian distributions with widths $\Delta_q$ and $\Delta_p$ respectively. If $\Delta_q$ and $\Delta_p$ are small, then in principle these shifts can be corrected with high probability: {\em e.g}, for $\Delta_q=\Delta_p\equiv \Delta$, the probability that a shift in $q$ or $p$ causes an uncorrectable error is no worse than the probability that the size of the shift exceeds $\sqrt{\pi}/2$, or
\begin{eqnarray}
{\rm Error ~ Prob} &\le &{2\over \sqrt{\pi\Delta^2}}\int_{\sqrt{\pi}/2}^\infty dq ~e^{-q^2/\Delta^2}\nonumber \\
&\le &{2\Delta\over \pi}~\exp(-\pi/4\Delta^2)~.
\end{eqnarray}
For the $d=2$ code with $\alpha\ne 1$, this same estimate of the error probability applies if we rescale the widths appropriately,
\begin{equation}
\Delta_q= \Delta\cdot \alpha~,\quad \Delta_p=\Delta/\alpha~.
\end{equation}

We can concatenate a shift-resistant code with an $[[n,k,d]]$ stabilizer quantum code. That is, first we encode (say) a qubit in each of $n$ oscillators; then $k$ better protected qubits are embedded in the block of $n$. If the typical shifts are small, then the qubit error rate will be small in each of the $n$ oscillators, and the error rate in the $k$ protected qubits will be much smaller. The quantum key distribution protocols that we propose are based on such concatenated codes. 

We note quantum codes for continuous quantum variables with an {\em infinite-dimensional} code space were described earlier by Braunstein \cite{braunstein}, and by Lloyd and Slotine \cite{lloyd}. Entanglement distillation protocols for continuous variable systems have also been proposed \cite{plenio,zoller}

\section{Quantum key distribution and quantum error-correcting codes}
\label{sec:qkd_qecc}

Now let's recall the connection between stabilizer quantum codes and quantum key distribution schemes \cite{shor_pres}.  

We say that a protocol for quantum key distribution is secure if (1) the eavesdropper Eve is unable to collect a significant amount of information about the key without being detected, (2) the communicating parties Alice and Bob receive the same key bits with high probability, and (3) the key generated is essentially random. Then if the key is intercepted, Alice and Bob will know it is unsafe to use the key and can make further attempts to establish a secure key. If eavesdropping is not detected, the key can be safely used as a one-time pad for encoding and decoding.\footnote{We implicitly assume that Eve uses a strategy that passes the verification test with nonnegligible probability, so that the rate of key generation is not exponentially small. If, for example, Eve were to intercept all qubits sent by Alice and resend them to Bob, then she would almost certainly be detected, and key bits would not be likely to be generated. But in the rare event that she is not detected and some key bits are generated, Eve would know a lot about them.}

Establishing that a protocol is secure is tricky, because there inevitably will be some noise in the quantum channel used to distribute the key, and the effects of eavesdropping could be confused with the effects of the noise. Hence the protocol must incorporate error correction to establish a shared key despite the noise, and privacy amplification to control the amount of information about the key that can be collected by the eavesdropper.

In the case of the BB84 key distribution invented by Bennett and Brassard \cite{bb84}, the necessary error correction and privacy amplification are entirely classical. Nevertheless, the formalism of quantum error correction can be usefully invoked to show that the error correction and privacy amplification work effectively \cite{shor_pres}. The key point is that if Alice and Bob carry out the BB84 protocol, we can show that the eavesdropper is no better off than if they had executed a protocol that applies quantum error correction to the transmitted quantum states. Appealing to the observation that Alice and Bob {\em could have} applied quantum error correction (even though they didn't really apply it), we place limits on what Eve can know about the key.

\subsection{Entanglement distillation}
First we will describe a key distribution protocol that uses a quantum error-correcting code to purify entanglement, and will explain why the protocol is secure. The connection between quantum error correction and entanglement purification was first emphasized by Bennett {\it et al.} \cite{bdsw}; our proof of security follows a proof by Lo and Chau \cite{chau_lo} for a similar key distribution protocol. Later, following \cite{shor_pres}, we will see how the entanglement-purification protocol is related to the BB84 protocol.

A stabilizer code can be used as the basis of an entanglement-purification protocol with one-way classical communication \cite{bdsw,chau_lo}. Two parties, both equipped with quantum computers, can use this protocol to extract from their initial shared supply of noisy Bell pairs a smaller number of Bell pairs with very high fidelity. These purified Bell pairs can then be employed for EPR quantum key distribution. Because the distilled pairs are very nearly pure, the quantum state of the pairs has negligible entanglement with the quantum state of the probe of any potential eavesdropper; therefore no measurement of the probe can reveal any useful information about the secret key.

Let's examine the distillation protocol in greater detail. Suppose that Alice and Bob start out with $n$ shared EPR pairs. Ideally, these pairs should be in the state 
\begin{equation}
|\Phi^{(n)}\rangle \equiv |\phi^+\rangle^{\otimes n}~,
\end{equation}
where $|\phi^+\rangle$ is the Bell state $(|00\rangle + |11\rangle)/\sqrt{2}$; however, the pairs are noisy, approximating $|\Phi^{(n)}\rangle$ with imperfect fidelity. They wish to extract $k < n$ pairs that are less noisy. 

For this purpose, they have agreed in advance to use a particular $[[n,k,d]]$ stabilizer code. The code space can be characterized as a simultaneous eigenspace of a set of mutually commuting stabilizer generators $\{M_i, i=1,2,\dots,n-k\}$.  Each $M_i$ is a ``Pauli operator,'' a tensor product of $n$ single-qubit operators where each single-qubit operator is one of $\{I,X,Y,Z\}$ defined by
\begin{eqnarray}
&I= \pmatrix{1&0\cr 0&1\cr},~&X= \pmatrix{0&1\cr 1&0\cr},\nonumber\\
~&Y= \pmatrix{0&-i\cr i&0\cr},~&Z= \pmatrix{1&0\cr 0&-1 \cr}.
\end{eqnarray}
The operations $\{\bar X_a,\bar Z_a, a=1,2,\dots, k\}$ acting on the encoded qubits are Pauli operators that commute with all of the  $M_i$.

The Bell state $|\phi^+\rangle$ is the simultaneous eigenstate with eigenvalue one of the two commuting operators $X_A\otimes X_B$ and $Z_A\otimes Z_B$ (where subscripts $A$ and $B$ indicate whether the operator acts on Alice's or Bob's qubit). Thus the state $|\Phi^{(n)}\rangle$ is the simultaneous eigenstate with eigenvalue one of the commuting operators
\begin{eqnarray}
& M_{i,A}\otimes M_{i,B}~, \quad&i=1,2,\dots, n-k~,\nonumber\\
&\bar X_{a,A}\otimes \bar X_{a,B}~, \quad &a=1,2,\dots, k~,\nonumber\\
&\bar Z_{a,A}\otimes \bar Z_{a,B}~, \quad &a=1,2,\dots, k~.
\end{eqnarray}
Now suppose that Alice and Bob both measure the $n-k$ commuting $M_i$'s. If the state they measure is precisely $|\Phi^{(n)}\rangle$, then Alice and Bob obtain identical measurement outcomes. Furthermore, since their measurements do not disturb the encoded operations $\bar X_a$ and $\bar Z_a$, their measurement would prepare the encoded state $|\bar \Phi^{(k)}\rangle\equiv |\bar \phi^+\rangle^{\otimes k}$, the encoded state with
\begin{eqnarray}
\bar X_{a,A}\otimes \bar X_{a,B}&=&\bar Z_{a,A}\otimes \bar Z_{a,B}=1~, \nonumber\\
& &\quad a=1,2,\dots, k~,
\end{eqnarray}
in the code subspace with the specified values of $M_i=\pm 1$.

However, since the initial pairs are noisy, Alice's and Bob's measurement of the $M_i$'s need not match perfectly; they should apply error correction to improve the fidelity of their encoded pairs. Thus Alice broadcasts the values of the $M_{i,A}$'s that she obtained in her measurements. Comparing to his own measurements, Bob computes the relative syndrome $M_{i,A}\cdot M_{i,B}$. From this relative syndrome, he infers what recovery operation he should apply to his qubits to ensure that the $M_{i,B}$'s match the $M_{i,A}$'s, and he performs this operation.
Now Alice and Bob are in possession of $k$ encoded pairs with improved fidelity. 

These encoded pairs can be used for EPR key distribution. For each $a=1,2,\dots,k$, Alice and Bob measure $\bar Z_a$, obtaining outcomes that are essentially random and agree with high probability. These outcomes are their shared private key. 

\subsection{Verification}

If the initial pairs are {\em too} noisy, either because of the intervention of an eavesdropper or for other reasons, then the purification protocol might not succeed. Alice and Bob need to sacrifice some of their EPR pairs to verify that purification is likely to work. If verification fails, they can abort the protocol.

Under what conditions will purification succeed? If their pairs were perfect, each would be in the state $|\phi^+\rangle$, the simultaneous eigenstate with eigenvalue one of the two commuting observables $X\otimes X$ and $Z\otimes Z$. Suppose for a moment, that each of the pairs {\em is} a simultaneous eigenstate of these observables (a Bell state), but not necessarily with the right eigenvalues: in fact no more than $t_X$ of the $n$ pairs have $X\otimes X=-1$, and no more than $t_Z$ of the $n$ pairs have $Z\otimes Z=-1$. Then, if Alice and Bob use a stabilizer code that can correct up to $t_Z$ bit flip errors and up to $t_X$ phase errors, the purification protocol will work perfectly --- it will yield the encoded state $|\bar\Phi^{(k)}\rangle=|\bar \phi^+\rangle^{\otimes k}$ with fidelity $F=1$.

Now, the initial $n$ pairs might not all be in Bell states. But suppose that Alice and Bob were able to perform a Bell measurement on each pair, projecting it onto a simultaneous eigenstate of $X\otimes X$ and $Z\otimes Z$. Of course, since Alice and Bob are far apart from one another, they cannot really do this Bell measurement. But let's nevertheless imagine that they first perform a Bell measurement on each pair, and then proceed with the purification protocol. Purification works if the Bell measurement yields no more than $t_X$ pairs with $X\otimes X=-1$ and no more than $t_Z$ pairs with $Z\otimes Z=-1$. Therefore, if the initial state of the $n$ pairs has the property that Bell measurement applied to all the pairs will, with very high probability, produce pairs with no more than $t_Z$ bit flip errors and no more than $t_X$ phase errors, then we are assured that Bell measurement followed by purification will produce a very high fidelity approximation to the encoded state $|\bar\Phi^{(k)}\rangle$.

But what if Alice and Bob execute the purification protocol without first performing the Bell measurement? We know that the purification works perfectly applied to the space ${\cal H}_{\rm good}$ spanned by Bell pairs that differ from $|\phi^+\rangle^{\otimes n}$ by no more than $t_Z$ bit flip errors and no more than $t_X$ phase errors. Let $\Pi$ denote the projection onto ${\cal H}_{\rm good}$ . Then if the protocol is applied to an initial density operator $\rho$ of the $n$ pairs, the final density operator $\rho'$ approximates $|\bar\Phi^{(k)}\rangle$ with fidelity
\begin{equation}
\label{fidelity_bound}
F\equiv \langle \bar\Phi^{(k)}|\rho'|\bar\Phi^{(k)}\rangle \ge {\rm tr}(\Pi\rho)~.
\end{equation}
Therefore, the fidelity is at least as large as the probability that $t_Z$ or fewer bit flip errors and $t_X$ or fewer phase errors would have been found if Bell measurement had been performed on all $n$ pairs.

To derive the inequality eq.~(\ref{fidelity_bound}), we represent $\rho$ as a pure state $|\Psi\rangle_{SE}$ of the $n$ pairs (the ``system'' $S$) and an ancilla (the ``environment'' $E$, which might be under Eve's control). The recovery superoperator can be represented as a unitary operator $U_{SR}$ that is applied to $S$ and an auxiliary system (the ``reservoir'' $R$) that serves as a repository for the entropy drawn from the pairs by error correction. Denote the initial pure state of the reservoir by $|0\rangle_R$. Then the state of system, environment, and reservoir to which the recovery operation is applied can be resolved into a ``good'' component
\begin{equation}
|\Psi_{\rm good}\rangle_{SER}= \left(\Pi_S\otimes I_{ER}\right)|\Psi\rangle_{SE}\otimes |0\rangle_R~,
\end{equation}
and an orthogonal component
\begin{equation}
|\Psi_{\rm bad}\rangle_{SER}= \left((I_S-\Pi_S)\otimes I_{ER}\right)|\Psi\rangle_{SE}\otimes |0\rangle_R~.
\end{equation}
Since the states $|\Psi_{\rm good}\rangle_{SER}$  and $|\Psi_{\rm bad}\rangle_{SER}$ are orthogonal, the unitary recovery operation $U_{SR}\otimes I_E$ maps them to states $|\Psi'_{\rm good}\rangle_{SER}$  and $|\Psi'_{\rm bad}\rangle_{SER}$ that are also orthogonal to one another.
Furthermore, since recovery works perfectly on the space ${\cal H}_{\rm good}$, we have
\begin{equation}
\label{good_junk}
|\Psi'_{\rm good}\rangle_{SER}= |\bar \Phi^{(k)}\rangle_S \otimes |{\rm junk}\rangle_{ER}~,
\end{equation}
where the state $|{\rm junk}\rangle_{ER}$ of environment and reservoir has norm
\begin{eqnarray}
& &{}_{ER}\langle{\rm junk}|{\rm junk}\rangle_{ER} = {}_{SER}\langle \Psi'_{\rm good}|\Psi'_{\rm good}\rangle_{SER}\nonumber\\
& & = {}_{SER}\langle \Psi_{\rm good}|\Psi_{\rm good}\rangle_{SER}= {\rm tr}(\Pi \rho)~.
\end{eqnarray}

Thus the fidelity of the recovered state can be expressed as
\begin{eqnarray}
\label{good_bad_fid}
F &=& {}_{SER}\langle \Psi'|\left(|\bar\Phi^{(k)}\rangle_S ~{}_S\langle \bar\Phi^{(k)}|\right)\otimes I_{ER} |\Psi'\rangle_{SER}\nonumber\\
&=& {}_{SER}\langle \Psi'_{\rm good}|\left(|\bar\Phi^{(k)}\rangle_S ~{}_S\langle \bar\Phi^{(k)}|\right)\otimes I_{ER} |\Psi'_{\rm good}\rangle_{SER}\nonumber\\
&+& {}_{SER}\langle \Psi'_{\rm bad}|\left(|\bar\Phi^{(k)}\rangle_S ~{}_S\langle \bar\Phi^{(k)}|\right)\otimes I_{ER} |\Psi'_{\rm bad}\rangle_{SER}\nonumber\\
&=& {\rm tr}(\Pi\rho) + \langle \bar\Phi^{(k)}|\rho'_{\rm bad}|\bar\Phi^{(k)}\rangle
\ge {\rm tr}(\Pi\rho)~ ,
\end{eqnarray}
where
\begin{equation}
\rho'_{\rm bad} = {\rm tr}_{ER} \left(|\Psi'_{\rm bad}\rangle_{SER}~{}_{SER}\langle \Psi'_{\rm bad}|\right)~;
\end{equation}
eq.~(\ref{fidelity_bound}) then follows. The key point is that, because of eq.~(\ref{good_junk}), and because $|\Psi'_{\rm good}\rangle_{SER}$  and $|\Psi'_{\rm bad}\rangle_{SER}$  are orthogonal, there is no ``good-bad'' cross term in eq.~(\ref{good_bad_fid}).

Our arguments so far show that Alice and Bob can be assured that entanglement purification will work very well if they know that it is highly unlikely that more than $t_Z$ bit flip errors or more than $t_X$ phase errors would have been found if they had projected their pairs onto the Bell basis. While they have no way of directly checking whether this condition is satisfied, they can conduct a test that, if successful, will provide them with high statistical confidence. We must now suppose that Alice and Bob start out with more than $n$ pairs; to be definite, suppose they have about $2n$ to start, and that they are willing to sacrifice about half of them to conduct their verification test. Alice randomly decides which pairs are for verification (the ``check pairs'') and which are for key distribution (the ``key pairs''), and  for each of her check qubits, she randomly decides to measure either $X$ or $Z$. Then Alice publicly announces which are the check pairs, whether she measured $X$ or $Z$ on her half of each check pair, and the results of those measurements (in addition to the results of her measurements of the stabilizer generators).

Upon hearing of Alice's choices, Bob measures $X$ or $Z$ on his half of each of the check pairs; thus Alice and Bob are able to measure $X\otimes X$ on about half of their check pairs, and they measure  $Z\otimes Z$ on the remaining check pairs.  Now since the check pairs were randomly chosen, the eavesdropper Eve has no way of knowing which are the check pairs, and she can't treat them any differently than the key pairs; hence the measured error rate found for the check pairs will be representative of the error rate that would have been found on the key pairs if Alice and Bob had projected the key pairs onto the Bell basis. Therefore, Alice and Bob can use their check data and classical sampling theory to estimate how many bit flip and phase errors would have been expected if they had measured the key pairs. 

For example, in a sample of $N$ pairs, suppose that {\em if} Alice and Bob both measured $Z$ for {\em all} the pairs, a fraction $p$ of their measurements would disagree, indicating bit flip errors. Then if they randomly sample $M<N$ of the pairs, the probability distribution for the number $M(p-\varepsilon)$ of errors found would be\footnote{This bound is not tight. It applies if the sample of $M$ pairs is chosen from the population of $N$ {\em with replacement}. In fact the sample is chosen without replacement, which suppresses the fluctuations. A better bound was quoted in \cite{shor_pres}.}
\begin{equation}
P(\varepsilon)< \exp(-M\varepsilon^2/2p(1-p))~.
\end{equation} 
If Alice and Bob have no {\em a priori} knowledge of the value of $p$, then by Bayes' theorem, the conditional probability that the total number of errors in the population is $pN$, given that there are $p_Z M$ errors in the sample, is the same as the probability that there are $p_Z M$ errors in the sample given that there are $pN$ errors in the total population. Writing $p=p_Z+\varepsilon$, the number of errors on the $N-M$ untested pairs is $Np - Mp_Z= (N-M)p_Z + N\varepsilon= (N-M)\cdot (p_Z + \varepsilon')$, where $\varepsilon'= N\varepsilon/(N-M)$. Expressing $P(\varepsilon)$ in terms of $\varepsilon'$ we find
\begin{equation}
P(\varepsilon')< \exp\left(- {M(N-M)^2\varepsilon'^2\over 2 N^2 p_Z(1-p_Z)}\right)~,
\end{equation}
a bound on the probability that the fraction of the untested pairs with errors is larger than $p_Z+\varepsilon'$. In particular, if they test about $M=n/2$ pairs for bit flip errors out of a total of about $N= n + n/2$ pairs, the probability that a fraction $p_Z + \varepsilon'$ of the remaining $N-M=n$ pairs have bit flip errors is 
\begin{equation}
P(\varepsilon')< \exp\left(- {n\varepsilon'^2/ 9 p_Z(1-p_Z)}\right)~.
\end{equation}
A similar argument applies to the probability of phase errors. We conclude that by conducting the verification test, Alice and Bob can be very confident that, if they had measured $Z\otimes Z$ (or $X\otimes X$) on the $n$ key pairs,  no more than $(p_Z+\varepsilon')n$ (or $(p_X+\varepsilon')n$) errors would have been found. By choosing a quantum error-correcting code that can correct this many errors with high probability, they can be confident that the encoded state they prepare approximates $|\bar\Phi^{(k)}\rangle$ with fidelity exponentially close to one. 

It is important to emphasize that this argument requires no assumption about how the errors on different pairs may be correlated with one another. Rather the argument is applied to a hypothetical situation in which the value of $Z\otimes Z$ (or $X\otimes X$) already has been measured and recorded for all of the check pairs and all of the key pairs. Sampling theory is then used to address the question: how reliably does a ``poll'' of $M$ bits randomly chosen from among $N$ allow us to predict the behavior of the rest of the population. Classical sampling theory can be applied to the values of both $Z\otimes Z$ and $X\otimes X$ for the key pairs, because the operators commute and so are simultaneously measurable in principle \cite{chau_lo}.

Furthermore, if the state of the encoded pairs that Alice and Bob use for key distribution is exponentially close to being a pure state, it follows from Holevo's theorem 
that Eve's mutual information with the distributed key is exponentially small \cite{chau_lo,shor_pres}. In the worst case, the imperfect fidelity of Alice's and Bob's pairs is entirely due to Eve's intervention; then the complete state consisting of the pairs and Eve's probe is pure, and the Von Neumann entropy $S(\rho_E)\equiv -{\rm tr}~\rho_E\log\rho_E$ of the state  $\rho_E$ of the probe equals the entropy of the state $\rho_{AB}$ of the pairs. By extracting a key from their pairs, Alice and Bob in effect prepare a state for Eve governed by an ensemble with density matrix $\rho_E$. According to Holevo's theorem, the mutual information $I(AB;E)$ of this state preparation with any measurement that Eve can carry out on her probe satisfies
\begin{equation}
I(AB;E)\le S(\rho_E)= S(\rho_{AB})~,
\end{equation}
and since $\rho_{AB}$ is very nearly pure, $S(\rho_{AB})$ and $I(AB;E)$ are very close to zero.
Specifically, if the fidelity of $\rho_{AB}$ is $F=1-\delta$, then the largest eigenvalue of $\rho_{AB}$ is at least $1-\delta$. For a system with dimension $D$, the density matrix with largest eigenvalue $1-\delta$ that has the maximal Von Neumann entropy is 
\begin{equation}
\rho_{\rm max}= {\rm diag}\left(1-\delta, {\delta\over D-1},{\delta\over D-1}, \dots, {\delta\over D-1}\right).
\end{equation}
for which
\begin{eqnarray}
&&S(\rho_{\rm max})= -(1-\delta)\log_2(1-\delta) - \delta\log_2(\delta/(D-1))\nonumber\\
&=& \delta\cdot\left( {1\over \log_e 2} + \log_2(D-1) -\log_2\delta\right) + O(\delta^2)~.
\end{eqnarray}
Taking $D=2^{2k}$ (the total dimension of Alice's and Bob's code spaces), we conclude that
\begin{equation}
S(\rho_{AB}) \le \delta\cdot \left( {1\over \log_e 2} + 2k +\log_2(1/\delta)\right) + O(\delta^2)~.
\end{equation}

Finally, we have shown that if the verification test succeeds, then with probability exponentially close to one (the probability that the error rate inferred from the check sample is not seriously misleading), Eve's mutual information with the key is exponentially small (because the state of the key bits approximates $|\bar\Phi^{(k)}\rangle$ with fidelity exponentially close to one). This proof of security applies to any conceivable eavesdropping strategy adopted by Eve. 

The proof relies on the ability of quantum error-correcting codes to reverse the errors caused by interactions between the key pairs and Eve's probe. Hence it may seem odd that the proof works for arbitrary attacks by Eve, since quantum error correction works effectively only for a restricted class of error superoperators. Specifically, the error superoperator acting on a block of $n$ qubits can be expanded in terms of a basis of ``Pauli error operators,'' where in each term of the expansion bit flip errors and/or phase errors are inflicted on specified qubits within the block.  The encoded quantum information is well protected only if the error superoperator has nearly all of its support on Pauli operators that can be corrected by the code, {\it e.g.}, those with no more than $t_Z$  bit flip errors and $t_X$ phase errors. 

If Eve's probe interacts collectively with many qubits, it may cause more bit flip or phase errors than the code can correct. But the crucial point is that, with high probability, an attack that causes many errors on the key bits will also cause many errors on the check bits, and Alice and Bob will detect Eve's presence.

\subsection{Reduction to the BB84 protocol}

Since the entanglement distillation protocol requires only one-way classical communication, this protocol is actually equivalent to one in which Alice, rather than preparing Bell pairs and sending half of each pair to Bob, instead prepares an encoded quantum state that she sends to Bob. Using a set of stabilizer generators on which she and Bob have agreed in advance, Alice chooses a random eigenvalue for each stabilizer generator $M_i$; then employing the corresponding $[[n,k,d]]$ quantum code, she prepares one of $2^k$ mutually orthogonal codewords.  

Alice also decides at random which of her qubits will be used for key distribution and which will be used for verification. For each of the check bits, she decides at random whether to send an $X$ eigenstate (with random eigenvalue) or a $Z$ eigenstate (with random eigenvalue). 

Bob receives the qubits sent by Alice, carefully deposits them in his quantum memory, and publicly announces that the qubits have been received. Alice then publicly reveals which qubits were used for the key, and which qubits are the check qubits. She announces the stabilizer eigenvalues that she chose to encode her state, and for each check qubit, she announces whether it was prepared as an $X$ or $Z$ eigenstate, and with what eigenvalue. 

Once Bob learns which qubits carry the encoded key information, he measures the stabilizer operators and compares his results with Alice's to obtain a relative error syndrome. He then performs error recovery and measures the encoded state to decipher the key. 

Bob also measures the check qubits and compares the outcomes to the values announced by Alice, to obtain an estimate of the error rate. If the error rate is low enough, error recovery applied to the encoded key bits will succeed with high probability, and Alice and Bob can be confident in the security of the key. If the error rate is too high, Bob informs Alice and they abort the protocol.

As described so far, the protocol requires that Alice and Bob have quantum memories and quantum computers that are used to store the qubits, measure stabilizer generators, and correct errors. But if they use a stabilizer code of the CSS (Calderbank-Shor-Steane) type \cite{calderbank,steane}, then the protocol can be simplified further. The crucial property of the CSS codes is that there is a clean separation between the syndrome information needed to correct bit flip errors and the syndrome information needed to correct phase errors.

A CSS quantum stabilizer code is associated with a classical binary linear code $C_1$ on $n$ bits, and a subcode $C_2\subset C_1$. Let $H_1$ denote the parity check matrix of $C_1$ and $H_2$ the generator matrix for the code $C_2$ (and hence the parity check matrix of the dual code $C_2^\perp$). The stabilizer generators of the code are of two types. Associated with the $i$th row of the matrix $H_1$ is a ``$Z$-generator,'' the tensor product of $I$'s and $Z$'s \begin{equation}
M_{Z,i}= \otimes_{j=1}^n (Z_j)^{(H_1)_{ij}}~,
\end{equation}
and associated with the $i$th row of $H_2$ is an ``$X$-generator,'' the tensor product of $I$'s and $X$'s
\begin{equation}
M_{X,i}= \otimes_{j=1}^n (X_j)^{(H_2)_{ij}}~.
\end{equation}
Since $H_1$ has $n- k_1$ rows, where $k_1={\rm dim}(C_1)$, and $H_2$ has $k_2$ rows, where $k_2 ={\rm dim}(C_2)$ there are all together $n-k_1+k_2$ stabilizer generators, and the dimension of the code space (the number of encoded qubits) is $k=k_1-k_2$. From measurements of the $Z$ generators, bit flip errors can be diagnosed, and from measurement of the $X$ generators, phase errors can be diagnosed.

The elements of a basis for the code space with eigenvalues of stabilizer generators
\begin{equation}
M_{Z,i}=(-1)^{s_i}~,\quad M_{X,i}=(-1)^{t_i}
\end{equation}
are in one-to-one correspondence with the $k$ cosets of $C_2$ in $C_1$; they can be chosen as
\begin{equation}
|\psi(v)\rangle_{x,z}= {1\over |C_2|^{1/2}}\sum_{w\in C_2}(-1)^{z\cdot w}|v+w+x\rangle~;
\end{equation}
here $v\in C_1$ is a representative of a $C_2$ coset, and $x$, $z$ are $n$-bit strings satisfying
\begin{equation}
H_1 x=s~,\quad H_2 z=t~.
\end{equation}
Thus, to distribute the key, Alice chooses $x$ and $z$ at random, encodes one of the $|\psi(v)\rangle_{x,z}$'s, and sends the state to Bob. After Bob confirms receipt, Alice broadcasts the values of $x$ and $z$.  Bob compares Alice's values to his own measurements of the stabilizer generators to infer a relative syndrome, and he performs error correction. Then Bob measures $Z$ of each of his $n$ qubits, obtaining a bit string $v+w+x$.  Finally, he subtracts $x$ and applies $H_2$ to compute $H_2v$, from which he can infer the coset represented by $v$ and hence the key.

Now notice that Bob extracts the encoded key information by measuring $Z$ of each of the qubits that Alice sends. Thus Bob can correctly decipher the key information by correcting any bit flip errors that occur during transmission. Bob does not need to correct phase errors, and therefore he has no use for the phase syndrome information; hence there is no need for Alice to send it.

Without in any way weakening the effectiveness of the protocol, Alice can prepare the encoded state $|\psi(v)\rangle_{x,z}$, but discard her value of $z$, rather then transmitting it; thus we can consider the state sent by Alice to be averaged over the value of $z$. Averaging over the phase $(-1)^{z\cdot w}$ destroys the coherence of the sum over $w\in C_2$ in $|\psi(v)\rangle_{x,z}$;  in effect, then, Alice is preparing $n$ qubits as $Z$ eigenstates, in the state  $|v+w+x\rangle$, sending the state to Bob, and later broadcasting the value of $x$. We can just as well say that Alice sends a random string $u$, and later broadcasts the value of $u+v$. Bob receives $u+e$ (where $e$ has support on the bits that flip due to errors) extracts $v+e$, corrects it to the nearest $C_1$ codeword, and infers the key, the coset $v+C_2$.

Alice and Bob can carry out this protocol even if Bob has no quantum memory. Alice decides at random to prepare her qubits as $X$ or $Z$ eigenstates, with random eigenvalues, and Bob decides at random to measure in the $X$ or $Z$ basis. After public discussion, Alice and Bob discard the results in the cases where they used different bases and retain the results where they used the same basis. Thus the protocol we have described is just the BB84 protocol invented by Bennett and Brassard \cite{bb84}, accompanied by classical error correction (adjusting $v+e$ to a $C_1$ codeword) and privacy amplification (extracting the coset $v+C_2$).

What error rate is acceptable?  In a random CSS code, about half of the $n-k$
generators correct bit flips, and about half correct phase flips.  Suppose that the verification test finds that bit flip errors ($Z_A\otimes Z_B=-1$) occur with probability $p_Z$ and phase errors ($X_A\otimes X_B=-1)$ occur with probability $p_X$. Classical coding theory shows that a random CSS code can correct the bit flips with high probability if the number of typical errors on $n$ bits is much smaller than the number of possible bit flip error syndromes, which holds provided that  
\begin{equation}
{n\choose np_Z} 2^{- (n-k)/2}
\sim 2^{nH_{2}(p_Z) -  (n-k)/2}\ll 1~,
\end{equation}
where $H_2(x) = -x\log_2 x - (1-x)\log_2(1-x)$ is the binary entropy function.
Similarly, the phase errors can be corrected with high probability provided the same relation holds with $p_Z$ replaced by $p_X$.
Therefore, asymptotically as $n\to\infty$, secure key bits can be extracted from transmitted key bits at any rate $R$ satisfying 
\begin{eqnarray}
R = \frac{k}{n} < 1 - 2 H_2 (p_Z)~,\nonumber\\
R = \frac{k}{n} < 1 - 2 H_2 (p_X)~.
\end{eqnarray}
This upper bound on $R$ crosses zero at $p_Z$ (or $p_X)=.1100$.
We conclude that secure key distribution is possible if  $p_{X,Z}<11\%$. 

The random coding argument applies if the errors in the key qubits are randomly distributed. To assure that this is so, we can direct Alice to perform a random permutation of the qubits before sending them to Bob. After Bob confirms receipt, Alice can broadcast the permutation she performed, and Bob can invert it.

Again, the essence of this argument is that the amount of information that an eavesdropper could acquire is limited by how successfully we could have carried out quantum error correction if we had chosen to -- and that this relation holds irrespective of whether we really implemented the quantum error correction or not.

Other proofs of the security of the BB84 protocol have been presented \cite{mayers,biham}, which don't make direct use of this connection with quantum error-correcting codes. However, these proofs do use classical error correction and privacy amplification, and they implicitly exploit the structure of the CSS codes.

\subsection{Imperfect sources}
\label{subsec:bias}

Our objective in this paper is to analyze the security of key distribution schemes that use systems described by continuous quantum variables. The analysis will follow the strategy we have just outlined, in which an entanglement-purification protocol is reduced to a protocol that does not require the distribution of entanglement. But first we need to discuss a more general version of the argument.

In the entanglement-purification protocol, whose reduction to the BB84 protocol we have just described, there is an implicit limitation on the eavesdropper's activity. We have assumed that Alice prepares perfect entangled pairs in the state $|\phi^+\rangle$, and then sends half of each pair to Bob. Eve has been permitted to tamper with the qubits that are sent to Bob in any way she chooses, but she has not been allowed any contact with Alice's qubits. Therefore, if we imagine that Alice measures her qubits before sending to Bob, we obtain a BB84 protocol in which Alice is equipped with a perfect source of polarized qubits. When she sends a $Z$ eigenstate, the decision to emit a $|0\rangle$ or a $|1\rangle$ is perfectly random, and the state emerges from her source with perfect fidelity. Similarly, when she sends an $X$ eigenstate, the decision to send $|\pm\rangle\equiv (|0\rangle \pm |1\rangle)/\sqrt{2}$ is perfectly random, and the state is prepared with perfect fidelity. Furthermore, Eve has no knowledge of what Alice's source does, other than what she is able to infer by probing the qubits as they travel to Bob.

Security can be maintained in a more general scenario. In the entanglement-purification protocol, we can allow Eve access to Alice's qubits. As long as Eve has no way of knowing which pairs Alice and Bob will select for their verification test, and no way of knowing whether the check pairs will be measured in the $Z$ or $X$ basis, then the protocol still works: eavesdropping can be detected irrespective of whether Eve probes Alice's qubits, Bob's qubits, or both.

Now if we imagine that Alice measures her qubits before sending to Bob, we obtain a BB84-like protocol in which Alice's source is imperfect and/or Eve is able to collect some information about how Alice's source behaves. Our proof that the BB84-like protocol is secure still works as before. However the proof applies only to a restricted type of source --- it must be possible to simulate Alice's source exactly by measuring half of a two-qubit state.

To be concrete, consider the following special case, which will suffice for our purposes: Alice has many identical copies of the two-qubit state $\rho_{AB}$. To prepare a ``$Z$-state'' she measures qubit $A$ in the basis $\{|0\rangle_A,|1\rangle_A\}$. Thus she sends to Bob one of the two states
\begin{eqnarray}
\rho_{0} &=& {{}_A\langle 0|\rho_{AB}|0\rangle_A\over  {\rm tr}\left({}_A\langle 0|\rho_{AB}|0\rangle_A\right)} ~,\nonumber\\
\rho_{1}&=& {{}_A\langle 1|\rho_{AB}|1\rangle_A\over  {\rm tr}\left({}_A\langle 1|\rho_{AB}|1\rangle_A\right)} ~,
\end{eqnarray}
chosen with respective probabilities
\begin{eqnarray}
{\rm Prob}(0) &=& {\rm tr}\left({}_A\langle 0|\rho_{AB}|0\rangle_A\right) ~,\nonumber\\
{\rm Prob}(1) &=& {\rm tr}\left({}_A\langle 1|\rho_{AB}|1\rangle_A\right) ~.
\end{eqnarray}
Similarly, to prepare an $X$-state she measures in the basis $\{|+\rangle,|-\rangle\}$, sending one of 
\begin{eqnarray}
\rho_{+} &=& {{}_A\langle +|\rho_{AB}|+\rangle_A\over  {\rm tr}\left({}_A\langle +|\rho_{AB}|+\rangle_A\right)} ~,\nonumber\\
\rho_{-}&=& {{}_A\langle -|\rho_{AB}|-\rangle_A\over  {\rm tr}\left({}_A\langle -|\rho_{AB}|-\rangle_A\right)} ~,
\end{eqnarray}
chosen with respective probabilities
\begin{eqnarray}
{\rm Prob}(+) &=& {\rm tr}\left({}_A\langle +|\rho_{AB}|+\rangle_A\right) ~,\nonumber\\
{\rm Prob}(-) &=& {\rm tr}\left({}_A\langle -|\rho_{AB}|-\rangle_A\right) ~.
\end{eqnarray}
Unless the state $\rho_{AB}$ is precisely the pure state $|\phi^+\rangle$, Alice's source isn't doing exactly what it is supposed to do. Depending on how $\rho_{AB}$ is chosen, the source might be biased; for example it might send $\rho_{0}$ with higher probability than $\rho_{1}$. And the states $\rho_{0}$ and $\rho_{1}$ need not be the perfectly prepared $|0\rangle$ and $|1\rangle$ that the protocol calls for.

Now suppose that Alice's source always emits one of the states $\rho_{0},\rho_1,\rho_+,\rho_-$, and that after the qubits emerge from the source, Eve is free to probe them any way she pleases. Even though Alice's source is flawed, Alice and Bob can perform verification, error correction, and privacy amplification just as in the BB84 protocol. To verify, Bob measures $Z$ or $X$, as before; if he measures $Z$, say, they check to see whether Bob's outcome $|0\rangle$ or $|1\rangle$ agrees with whether Alice sent $\rho_0$ or $\rho_1$ (even though the state that Alice sent may not have been a $Z$ eigenstate). Thereby, Alice and Bob estimate error rates $p_Z$ and $p_X$. If both error rates are below $11\%$, then the protocol is secure.

We emphasize again that the security criterion $p_{X},p_{Z} < 11 \%$ applies not to all sources, but only to the restricted class of imperfect sources that can be simulated by measuring half of a (possible noisy) entangled state. To give an extreme example of a type of source to which the security proof does not apply, suppose that Alice {\em always} sends the $Z$-state $|0\rangle$ or  the $X$-state $|+\rangle$. Clearly the key distribution protocol will fail, even if Bob's bits always agree with Alice's! Indeed, a source with these properties cannot be obtained by measuring half of any two-qubit state $\rho_{AB}$. Rather, if the source is obtained by such a measurement, then a heavy bias when we send a $Z$-state would require that the error probability be large when we send an $X$-state.

\section{Distributing a key bit with continuous variables}
\label{sec:qkd_cont}

Now let's consider how the above ideas can be applied to continuous variable systems. We will first describe how in principle Alice and Bob can extract good encoded pairs of qubits from noisy EPR pairs. However, the distillation protocol requires them to make measurements that are difficult in practice. Then we will see how key distribution that invokes (difficult) entanglement distillation can be reduced to key distribution based on (easier) preparation, transmission, and detection of squeezed states. 

Suppose that Alice and Bob share pairs of oscillators. Ideally each pair has been prepared in an EPR state, a simultaneous eigenstate (let's say with eigenvalue 0) of $q_A-q_B$ and $p_A+p_B$. Now suppose that Alice measures the two commuting stabilizer generators defined in eq.~(\ref{stabilizer}), obtaining the outcomes
\begin{equation}
S_{q,A}=e^{2\pi i\phi_{q,A}}~,\quad S_{p,A}= e^{-2\pi i\phi_{p,A}}~,
\end{equation}
or 
\begin{eqnarray}
q_A &=& \phi_{q,A}\cdot \sqrt{\pi} ~({\rm mod}~\sqrt{\pi})~,\nonumber\\
p_A &=& \phi_{p,A}\cdot \sqrt{\pi} ~({\rm mod}~\sqrt{\pi})~.
\end{eqnarray}
Now, the initial state was an eigenstate with eigenvalue one of the operators $S_{q,A}\otimes S_{q,B}^{-1}$ and $S_{p,A}\otimes S_{p,B}$. The observables measured by Alice commute with these, and so preserve their eigenvalues. Thus if the initial EPR state of the oscillators were perfect, Alice's measurement would also prepare for Bob a simultaneous eigenstate of the stabilizer generators with
\begin{eqnarray}
S_{q,B}&\equiv& e^{2\pi i\phi_{q,B}}=e^{2\pi i\phi_{q,A}}~,\nonumber\\
 S_{p,B}&\equiv& e^{-2\pi i\phi_{p,B}}=e^{2\pi i\phi_{p,A}}~,
\end{eqnarray}
or
\begin{eqnarray}
q_B&=& q_A ~({\rm mod}~\sqrt{\pi})~,\nonumber\\
p_B&=& -p_A ~({\rm mod}~\sqrt{\pi})~.
\end{eqnarray}
Similarly, the initial state was an eigenstate with eigenvalue one of the observables
\begin{equation}
\bar X_A(\phi_p)\otimes \bar X_B(\phi_p)~,\quad \bar Z_A(\phi_q)\otimes\bar Z_B(\phi_q)^{-1}~,
\end{equation}
which also commute with the stabilizer generators that Alice measured. Thus Alice's measurement has prepared an encoded Bell pair in the code space labeled by $(\phi_q,\phi_p)$, the state
\begin{equation}
|\bar \phi^+\rangle_{AB} = {1\over \sqrt{2}}\left(|\bar 0\rangle_A |\bar 0\rangle_B + |\bar 1\rangle_A |\bar 1\rangle_B\right)~.
\end{equation}

Of course the initial EPR pair shared by Alice and Bob might be imperfect, and then the encoded state produced by Alice's measurement will also have errors. But if the EPR pair is not too noisy, they can correct the errors with high probability. Alice broadcasts her measured values of the stabilizer generators to Bob; Bob also measures the stabilizer generators and compares his values to those reported by Alice, obtaining a relative syndrome
\begin{equation}
e^{i(\phi_{q,A}-\phi_{q,B})}~,\quad  e^{-i(\phi_{p,A}+\phi_{p,B})}~.
\end{equation}
That is, the relative syndrome determines the value of $q_A-q_B$ (mod $\sqrt{\pi}$), and $p_A+p_B$ (mod $\sqrt{\pi}$). Using this information, Bob can shift his oscillator's $q$ and $p$ (by an amount between $-\sqrt{\pi}/2$ and $\sqrt{\pi}/2$) to adjust $q_A-q_B$ (mod $\sqrt{\pi}$), and $p_A+p_B$ (mod $\sqrt{\pi}$) both to zero. The result is that Alice and Bob now share a bipartite state in the code subspace labeled by $(\phi_q,\phi_p)$.

If the initial noisy EPR state differs from the ideal EPR state only by relative shifts of Bob's oscillator relative to Alice's that satisfy $|\Delta q|,|\Delta p|< \sqrt{\pi}/2$, then the shifts will be corrected perfectly. And if larger shifts are highly unlikely, then Alice and Bob will obtain a state that approximates the desired encoded Bell pair $|\bar \phi^+\rangle$ with good fidelity.  This procedure is a ``distillation'' protocol in that Alice and Bob start out with a noisy entangled state in a tensor product of infinite dimensional Hilbert spaces, and ``distill'' from it a far cleaner entangled state in a tensor product of two-dimensional subspaces.

Once Alice and Bob have distilled an encoded Bell pair, they can use it to generate a key bit, via the usual EPR key distribution protocol: Alice decides at random to measure either $\bar X$ or $\bar Z$, and then publicly reveals what she chose to measure but not the measurement outcome. Bob then measures the same observable and obtains the same outcome -- that outcome is the shared key bit.

How do they measure $\bar X$ or $\bar Z$? If Alice (say) wishes to measure $\bar Z$, she can measure $q$, and then subtract $\phi_q$ from the outcome. The value of $\bar Z$ is determined by whether the result is an even ($\bar Z=1$) or an odd ($\bar Z=-1$) multiple of $\sqrt{\pi}$. Similarly, if Alice wants to measure $\bar X$, she measures $p$ and subtracts $\phi_p$ -- The value of $\bar X$ is determined by whether the result is an even ($\bar X=1$) or an odd ($\bar X=-1$) multiple of $\sqrt{\pi}$. 

Imperfections in the initial EPR pairs are inescapable not just because of experimental realities, but also because the ideal EPR pairs are unphysical nonnormalizable states. Likewise, the stabilizer operators cannot even in principle be measured with arbitrary precision (the result would be an infinite bit string), but only to some finite $m$-bit accuracy. Still, if the EPR pairs have reasonably good fidelity, and the measurements have reasonably good resolution, entanglement purification will be successful.

To summarize, Alice and Bob can generate a shared bit by using the continuous variable code for entanglement purification, carrying out this protocol:

\vspace{\baselineskip}

\centerline{{\bf Key distribution with entanglement purification}}
\begin{itemize}
\setlength{\itemsep}{-\parskip}

\item [\bf 1:] Alice prepares  (a good approximation to) an EPR state of two oscillators, a simultaneous eigenstate of $q_A-q_B= 0 = p_A + p_B$, and sends one of the oscillators to Bob.

\item [\bf 2:] After Bob confirms receipt, Alice and Bob each measure (to $m$ bits of accuracy) the two commuting stabilizer generators of the code, $e^{i(2\sqrt{\pi})q}$ and $e^{-i(2\sqrt{\pi})p}$. (Equivalently, they each measure the value of $q$ and $p$ modulo $\sqrt{\pi}$.) Alice broadcasts her result to Bob, and Bob applies shifts in $q$ and $p$ to his oscillator, so that his values of $q$ and $p$  modulo $\sqrt{\pi}$ now agree with Alice's (to $m$-bit accuracy). Thus, Alice and Bob have prepared (a very good approximation to) a Bell state $|\bar \phi^+\rangle$ of two qubits encoded in one of the simultaneous eigenspaces of the two stabilizer operators.

\item [\bf 3:] Alice decides at random to measure one of the encoded operators $\bar X$ or $\bar Z$; then she announces what she chose to measure, but not the outcome. Bob measures the same observable; the result is the shared bit that they have generated.

\end{itemize}

Now notice that, except for Bob's confirmation that he received the states, this protocol requires only one-way classical communication from Alice to Bob. Alice does not need to receive any information from Bob before she measures her stabilizer operators or before she measures the encoded operation $\bar X$ or $\bar Z$. Therefore, the protocol works just as well if Alice measures her oscillator before sending the other one to Bob. Equivalently, she prepares an encoded state, adopting randomly selected values of the stabilizer generators. She also decides at random whether the encoded state will be an $\bar X$ eigenstate or a $\bar Z$ eigenstate, and whether the eigenvalue will be $+1$ or $-1$. 

Again, since the codewords are unphysical nonnormalizable states, Alice can't really prepare a perfectly encoded state; she must settle for a ``good enough'' approximate codeword. 

In summary, we can replace the entanglement-purification protocol with this equivalent protocol:

\vspace{\baselineskip}

\centerline{{\bf Key distribution with encoded qubits}}
\begin{itemize}
\setlength{\itemsep}{-\parskip}

\item [\bf 1:] Alice chooses random values (to $m$ bits of accuracy) for the stabilizer generators $e^{i(2\sqrt{\pi})q}$ and $e^{-i(2\sqrt{\pi})p}$, chooses a random bit to decide whether to encode a $\bar Z$ eigenstate or an $\bar X$ eigenstate, and chooses another random bit to decide whether the eigenvalue will be $\pm 1$. She then prepares (a good approximation to) the encoded eigenstate of the chosen operator with the chosen eigenvalue in the chosen code, and sends it to Bob. 

\item [\bf 2:] After Bob confirms receipt, Alice broadcasts the stabilizer eigenvalues and whether she encoded a $\bar Z$ or an $\bar X$.

\item [\bf 3:] Bob measures $q$ or $p$.  He subtracts from his outcome the value modulo $\sqrt{\pi}$ determined by Alice's announced value of the stabilizer generator, and corrects the result to the nearest integer multiple of $\sqrt{\pi}$. He extracts a bit determined by whether the multiple of $\sqrt{\pi}$ is even or odd; this is the shared bit that they have generated.  

\end{itemize}

To carry out this protocol, Alice requires sophisticated tools that enable her to prepare the approximate codewords, and Bob needs a quantum memory to store the state that he receives until he hears Alice's classical broadcast. However, we can reduce the protocol to one that is much less technically demanding.

When Bob extracts the key bit by measuring (say) $q$, he needs Alice's value of $q$ modulo $\sqrt{\pi}$, but he does not need her value of the other stabilizer generator. Therefore, there is no need for Alice to send it; surely, the eavesdropper will be no better off if Alice sends less classical information. If she doesn't send the value of $S_p$, then we can consider the protocol averaged over the unknown value of this generator. Formally, for perfect (nonnormalizable) codewords the density matrix describing the state that is accessible to a potential eavesdropper then has a definite value of $S_q$ but is averaged over all possible values of $S_p$ -- it is a (nonnormalizable) equally weighted superposition of all position eigenstates with a specified value of $q$ mod $\sqrt{\pi}$; {\em e.g.} in the case where Alice prepares a $\bar Z$ eigenstate, we have
\begin{eqnarray}
& &\rho(\phi_q,\bar Z=1) \nonumber\\
& &\propto \sum_s |q=(2s+\phi_q)\sqrt{\pi}\rangle\langle q=(2s+\phi_q)\sqrt{\pi}|~,\nonumber\\
& &\rho(\phi_q,\bar Z=-1)\nonumber\\
& &\propto \sum_s |q=(2s+ 1+\phi_q)\sqrt{\pi}\rangle\langle q=(2s+1+\phi_q)\sqrt{\pi}|~.\nonumber\\
\end{eqnarray}
Averaged over $\phi_q$ as well, Alice is sending a random position eigenstate. Likewise, in the case where Alice prepares an $\bar X$ eigenstate, she sends a random momentum eigenstate. 

Therefore, the protocol in which Alice prepares encoded qubits can be replaced by a protocol that is simpler to execute but is no less effective and no less secure.  Instead of bothering to prepare the encoded qubit, she just decides at random to send either a $q$ or $p$ eigenstate, with a random eigenvalue. If Bob had a quantum memory, he could store the state, and wait to hear from Alice whether the state she sent was a $q$ or $p$ eigenstate; then he could measure that observable. Subtracting $\phi_q\sqrt{\pi}$ (or $\phi_p\sqrt{\pi}$) from his measurement outcome, he would obtain an even or odd multiple of $\sqrt{\pi}$. 

But Bob does not really need the quantum memory. As in the BB84 protocol, it suffices for Bob to decide at random to measure either $q$ or $p$, and then publicly compare his basis with Alice's. They discard the results where they used different bases and retain the others. 

A problem with this procedure is that the position and momentum eigenstates are unphysical nonnormalizable states, and the probability distribution that Alice samples to decide on what value of $q$ or $p$ to send is also nonnormalizable. For it to implementable, we need to modify the procedure so that Alice sends narrow $q$ or $p$ wave packets, and chooses the position of the center of the wave packet by sampling a broad but normalizable distribution. 

Therefore, Alice and Bob can adopt the following protocol:

\vspace{\baselineskip}

\centerline{{\bf Key distribution with squeezed states}}
\begin{itemize}
\setlength{\itemsep}{-\parskip}

\item [\bf 1:] Alice chooses a random bit to decide whether to send a state squeezed in $q$ or in $p$. She samples a (discrete approximation to) a probability distribution $P_{\rm pos}(q)$ or $P_{\rm mom}(p)$ to choose a value of $q$ or $p$, and then sends to Bob a narrow wave packet centered at that value. 

\item [\bf 2:] Bob receives the state and decides at random to measure either $q$ or $p$. 

\item [\bf 3:] After Bob confirms receipt, Alice and Bob broadcast whether they sent/measured in the $q$ or $p$ basis. If they used different bases, they discard their results. If they used the same basis, they retain the result and proceed to Step 4.

\item [\bf 4:] Alice broadcasts the value that she sent, modulo $\sqrt{\pi}$ (to $m$-bit accuracy). Bob subtracts Alice's value from what he measured, and corrects to the nearest integer multiple of $\sqrt{\pi}$. He and Alice extract their shared bit according to whether this integer is even or odd. 

\end{itemize}

\section{A secure protocol using continuous variables}
\label{sec:secure}
Now we are ready to combine the protocol of \S\ref{sec:qkd_qecc} with the protocol of \S\ref{sec:qkd_cont}. The result is a protocol based on concatenating the continuous variable code with an $[[n,k,d]]$ binary CSS code. The concatenated code embeds a $k$-dimensional Hilbert space in the infinite-dimensional Hilbert space of $n$ oscillators.

Again, we first imagine that Alice and Bob carry out an entanglement distillation protocol. They start out sharing $n$ pairs of oscillators, each in a (noisy) EPR state. By measuring the stabilizer generators of the concatenated code, they distill $k$ encoded Bell pairs of much better fidelity, and then generate a key by measuring the encoded Bell pairs.

By once again following the chain of reductions recounted in \S\ref{sec:qkd_qecc} and \S\ref{sec:qkd_cont}, we arrive at an equivalent protocol involving transmission of squeezed states. 
The complete protocol, including verification, error correction, and privacy amplification, becomes:

\vspace{\baselineskip}

\centerline{{\bf Continuous-variable QKD}}
\begin{itemize}
\setlength{\itemsep}{-\parskip}

\item [\bf 1:] Alice has $(4+\delta)n$  oscillators.  For 
	each oscillator, Alice decides at random to prepare either a state squeezed in $q$ or a state squeezed in $p$. The position of the squeezed state is determined by sampling (a discrete approximation to) a probability distribution $P_{\rm pos}(q)$ or $P_{\rm mom}(p)$. Alice then sends the oscillators to Bob.

\item [\bf 2:] Bob receives the $(4+\delta)n$ oscillators, measuring each
	in the $q$ or $p$ basis at random.

\item [\bf 3:] Bob confirms that the oscillators have been received, and then Alice announces whether each oscillator was squeezed in $q$ or in $p$.

\item [\bf 4:] Alice and Bob discard the results in the cases where Bob measured in a different
	basis than Alice used in her preparation.  With high probability, there are at
	least $2n$ measured values left (if not, abort the protocol).  Alice 
        decides randomly on a set of $2n$ values to use for the protocol, and
        chooses at random $n$ of these to be check values. 

\item [\bf 5:] For all $2n$ measured values, Alice announces the value of $q$ or $p$ modulo $\sqrt{\pi}$ (to $m$ bits of accuracy). 

\item [\bf 6:] Bob subtracts the corresponding number announced by Alice from each of his measured values, and then corrects the result to the nearest integer multiple of $\sqrt{\pi}$.  Bob and Alice now extract bit values determined by whether the multiple of $\sqrt{\pi}$ is even or odd.

\item [\bf 7:] Alice and Bob announce the values of their check bits.
	If too few of the check bits agree, they abort the protocol.

\item [\bf 8:] Alice announces $u+v$, where $u$ is the string
        consisting of the remaining non-check bits, and $v$ is a random 
        codeword in $C_1$.

\item [\bf 9:] Bob subtracts $u+v$ from his code qubits, $u+e$, and
        corrects the result, $v+e$, to a codeword in $C_1$.  With high probability, Bob recovers 
	$v$.

\item [\bf 10:] Alice and Bob use the $C_2$ coset $v + C_2$ as the key.

\end{itemize}

Here, to be specific, we have instructed Alice and Bob to sacrifice $n$ check bits for each $n$ bits that are used for key distribution. They might instead use fewer or more, depending on how stringent a bound on the eavesdropper's mutual information they require.

The check bits provide Alice and Bob with estimates of the bit error rates $p_Z$ (respectively $p_X$) when states squeezed in $q$ (respectively $p$) are transmitted. Our analysis of the BB84 protocol indicates that the squeezed state protocol is secure provided that  $p_Z$ and $p_X$ are both below $11\%$, and assuming that Alice and Bob scramble and unscramble the oscillators (by applying a random permutation and its inverse).

However, as noted in \S\ref{subsec:bias}, the proof and the security criterion $p_Z,p_X< 11\%$ apply only if Alice's source can be simulated by measuring half of an entangled state of two oscillators. In particular, we may imagine that Alice has many pairs of oscillators identically prepared in the state $\rho_{AB}$, and that she prepares the state that she sends to Bob by measuring oscillator $A$. When she measures in the $q$ basis, she sends the state
\begin{equation}
\rho_{B}(q)= {{}_A\langle q|\rho_{AB}|q\rangle_A\over
{\rm tr}\left({}_A\langle q|\rho_{AB}|q\rangle_A\right)}
\end{equation}
with probability 
\begin{equation}
P_{\rm pos}(q) = {\rm tr}\left({}_A\langle q|\rho_{AB}|q\rangle_A\right)~,
\end{equation}
and when she measures in the $p$ basis, she sends the state 
\begin{equation}
\rho_{B}(p)= {{}_A\langle p|\rho_{AB}|p\rangle_A\over
{\rm tr}\left({}_A\langle q|\rho_{AB}|q\rangle_A\right)}
\end{equation}
with probability 
\begin{equation}
P_{\rm mom}(p) = {\rm tr}\left({}_A\langle p|\rho_{AB}|p\rangle_A\right)~.
\end{equation}
Thus, the states that Alice sends need not be perfect position or momentum eigenstates for the proof of security to work, and Alice's source might even have a bias so that the raw key bit carried by an oscillator is more likely to be a 0 than a 1. Still, for a source of this type, if Alice and Bob verify that the error rate for the raw key bits is below $11\%$ in both bases, then the protocol is provably secure. We will discuss examples in \S\ref{sec:gaussian} and\S\ref{sec:losses}.

\begin{figure}
\begin{center}
\leavevmode
\epsfxsize=3in
\epsfbox{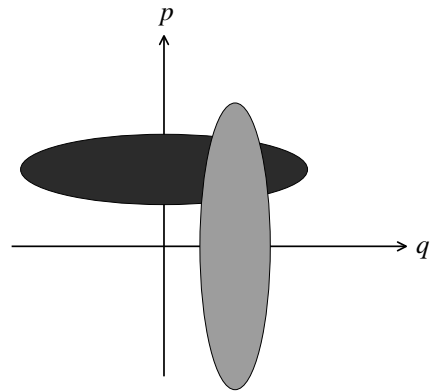}
\end{center}
\caption{One-sigma contours of the Wigner functions for typical squeezed states used in the quantum key distribution protocol, with squeeze factor $\tilde\Delta=e^{-r}=1/2$. The signal states squeezed in $p$ and in $q$ overlap with one another, preventing Eve from learning about one without disturbing the other.}
\label{fig:wigner}
\end{figure}

Intuitively, the squeezed state protocol is secure because the eavesdropper cannot monitor the value of $q$ (or $p$) transmitted without introducing a detectable disturbance in the complementary observable $p$ (or $q$). As shown in Fig.~\ref{fig:wigner}, the Wigner functions of the signal states squeezed in $p$ and in $q$ overlap, so that the states cannot be reliably distinguished.

\section{Gaussian states}
\label{sec:gaussian}

Perfectly squeezed states (position or momentum eigenstates) are unphysical nonnormalizable states, so the protocol will actually be carried out with imperfectly squeezed states. Furthermore, engineering a source that produces highly squeezed states would be quite technically demanding. How much squeezing is really needed for the protocol to be secure? A related question is, how must we choose the probability distributions $P_{\rm pos}(q)$ and $P_{\rm mom}(p)$ that govern the center of the squeezed state?

We will analyze the most favorable case, in which the squeezed states are Gaussian wave packets and the probability distributions are also Gaussian.  We will begin again with a description of how the code is used for entanglement purification, but where Alice and Bob start with many copies of a Gaussian entangled pair of oscillators that is an approximate eigenstate of $q_A-q_B$ and $p_A+p_B$. If we imagine that Alice measures half of each pair before she sends the other half to Bob, then we obtain a protocol in which Alice sends imperfectly squeezed states governed by a particular probability distribution.

The initial Gaussian entangled state of the two oscillators is
\begin{eqnarray}
\label{psi_delta}
|\psi(\Delta)\rangle_{AB} & = & {1\over \sqrt{\pi}}\int dq_A dq_B  ~\exp \left[- {1\over 2}\Delta^2\left({q_A+q_B\over 2}\right)^2\right] \nonumber\\
& &\times \exp\left[ - {1\over 2}\left(q_A-q_B\right)^2/\Delta^2\right]|q_A,q_B\rangle \nonumber\\
& =  &{1\over \sqrt{\pi}}\int dp_A dp_B  ~\exp \left[- {1\over 2}\Delta^2\left({p_A-p_B\over 2}\right)^2\right] \nonumber\\
&  &\times \exp\left[ - {1\over 2}\left(p_A+p_B\right)^2/\Delta^2\right]|p_A,p_B\rangle ~,
\end{eqnarray}
where $\Delta^2$ is real and positive. Since $|\psi(\Delta)\rangle_{AB}$ is actually invariant under 
\begin{equation}
\Delta^2\to 4/\Delta^2~, \quad q_B\to -q_B~, \quad p_B\to -p_B~
\end{equation}
we may assume without loss of generality (changing the sign of the position and momentum of Bob's oscillator if necessary), that $0< \Delta^2\le 2$. In the limiting case $\Delta^2=2$, $|\psi(\Delta)\rangle_{AB}$ becomes the product of two oscillator vacuum states. For $\Delta^2<2$, it is an entangled state. The amount of entanglement shared between the oscillators, in ``ebits,'' is defined as
\begin{equation}
E(\Delta)\equiv S(\rho_A)= -{\rm tr}~\rho_A\log_2 \rho_A~,
\end{equation}
(the Von Neumann entropy of Alice's density matrix $\rho_A = {\rm tr}_B  |\psi(\Delta)\rangle\langle \psi(\Delta)|)$, and can be expressed as \cite{vanenk}
\begin{eqnarray}
\label{ebits}
E(\Delta)= & &(\cosh^2 r) \log_2(\cosh^2 r) \nonumber\\
& - & (\sinh^2 r)\log_2(\sinh^2 r)~,
\end{eqnarray}
where 
\begin{equation}
\label{squeeze_param}
\Delta^2\equiv 2 e^{-2r}~.
\end{equation}

In this entangled state, if Alice measures the position of her oscillator and obtains the outcome $q_A$, she prepares for Bob the Gaussian state
\begin{eqnarray}
\label{psi_qa}
|\psi(q_A)\rangle_B &=& {1\over (\pi \tilde \Delta^2)^{1/4}}\int dq_B \nonumber\\
& \times & \exp\left(-{1\over 2}(q_B-q_{B0})^2/\tilde \Delta^2\right) |q_B\rangle~,
\end{eqnarray}
where 
\begin{equation}
q_{B0}= \left({{1 - {1\over 4} \Delta^4}\over{ 1+ {1\over 4}\Delta^4}}\right) ~q_A
=\left( 1 - \tilde \Delta^4\right)^{1/2}q_A~,
\end{equation}
and 
\begin{equation}
\label{tilde_delta}
\tilde \Delta^2= {\Delta^2\over 1+ {1\over 4}\Delta^4}~.
\end{equation}
The probability distribution for the outcome of Alice's measurement can be expressed as
\begin{equation}
\label{pqa}
P(q_A)={\tilde \Delta\over \sqrt{\pi}}~ \exp\left(-\tilde \Delta^2 q_A^2\right) ~,
\end{equation}
and we can easily see from eq.~(\ref{psi_delta}) that if Alice and Bob both measure $q$, then the difference of their outcomes is governed by the probability distribution
\begin{equation}
{\rm Prob}(q_A-q_B)= {1\over \sqrt {\pi \Delta^2}}\exp[ - (q_A-q_B)^2/\Delta^2]~.
\end{equation}
Similar formulas apply if Alice and Bob measure $p$.

Suppose that Alice and Bob try to distill one good qubit from the imperfect entangled state $|\psi(\Delta)\rangle_{AB}$. They both measure the stabilizer generators, that is, the values of $q$ and $p$ modulo $\sqrt{\pi}$. Alice broadcasts her values, and Bob adjusts his values so that they agree with Alice's; thereby they obtain a pair of encoded qubits, which would have been in the state $|\bar \phi^+\rangle$ if the initial pair of oscillators had been a perfect EPR pair ($\Delta^2=0$). Then if Alice and Bob were to proceed to perform a complete Bell measurement on their encoded qubit pair, the probability $p_Z$ that they would find $\bar Z\otimes \bar Z=-1$ is no worse than the probability that, if $q_A$ and $q_B$ were measured, the results would differ by more than $\sqrt{\pi}/2$, or 
\begin{eqnarray}
p_Z &\le &{2\over \sqrt{\pi\Delta^2}}\int_{\sqrt{\pi}/2}^\infty dq ~e^{-q^2/\Delta^2}\nonumber \\
&\le &{2\Delta\over \pi}~\exp(-\pi/4\Delta^2)~,
\end{eqnarray}
and similarly for $p_X$ (the probability that $\bar X\otimes \bar X=-1$). 
For the values of $\Delta$ that are typically of interest ({\em e.g.} $\Delta < 1$), the error probability is dominated by values of $q_A-q_B$ (or $p_A+p_B$) lying in the range $[\sqrt{\pi}/2,3\sqrt{\pi}/2]$, so that the estimate of the error probability can be sharpened to
\begin{equation}
\label{better_error}
p_Z,p_X \sim {2\over \sqrt{\pi\Delta^2}}\int_{\sqrt{\pi}/2}^{3\sqrt{\pi}/2} dq ~e^{-q^2/\Delta^2}~.
\end{equation}
After error correction and measurement in the encoded Bell basis, the initial bipartite pure state of two oscillators, with entanglement $E$ given by eq.~(\ref{ebits}) and (\ref{squeeze_param}), is reduced to a bipartite mixed state, diagonal in the encoded Bell basis, with fidelity $F=(1-p_Z)(1-p_X)$; this encoded state has entanglement of formation \cite{bdsw} 
\begin{equation}
\label{ent_form}
E=H_2\left({1\over 2} +\sqrt{F(1-F)}\right)
\end{equation}
(where $H_2$ is the binary entropy function). 

If Alice and Bob have a large number $n$ of oscillators in the state $|\psi(\Delta)\rangle_{AB}$, they can carry out an entanglement distillation protocol based on the concatenation of the single-oscillator code with a binary CSS code, and they will be able to distill qubits of arbitrarily good fidelity at a finite asymptotic rate provided that $p_Z$ and $p_X$ are both below $11\%$; from eq.~(\ref{better_error}) we find that this condition is satisfied for $\Delta < .784$ (which should be compared with the value $\Delta=\sqrt{2}$ corresponding to a product of two oscillators each in its vacuum state). Thus  secure EPR key distribution is possible in principle with two-mode squeezed states provided that the squeeze parameter $r$ satisfies $r > -\log_e (.784/\sqrt{2})= .590$; from eq.~(\ref{ebits}) and (\ref{ent_form}), $\Delta=.784$ corresponds to $E=1.19$ ebits carried by each oscillator pair,  which is reduced by error correction and encoded Bell measurement to $E=.450$ ebits carried by each of the encoded Bell pairs. 

Now consider the reduction of this entanglement distillation protocol to a protocol in which Alice prepares a squeezed state and sends it to Bob.  In the squeezed-state scheme, Alice sends the state $|\psi(q_A)\rangle$ with probability $P(q_A)$. The width $\tilde \Delta$ of the state that Alice sends is related to the parameter $\Delta$ appearing in the estimated error probability according to
\begin{equation}
\label{delta_tilde_delta}
\Delta^{-2}= \tilde\Delta^{-2}\cdot {1\over 2}(1 +\sqrt{1-\tilde\Delta^4})~.
\end{equation}
The state Alice sends is centered not at $q_A$ but at $q_{B0}= q_A\cdot (1-\tilde \Delta^4)^{1/2}$. Nevertheless, in the squeezed state protocol that we obtain as a reduction of the entanglement distillation protocol, it is $q_A$ rather than $q_{B0}$ that Alice uses to extract a key bit, and whose value modulo $\sqrt{\pi}$ she reports to Bob. The error probability that is required to be below $11\%$ to ensure security is the probability that error correction adjusts Bob's measurement outcome to a value that differs from $q_A$ (not $q_{B0}$) by an odd multiple of $\sqrt{\pi}$. As we have noted, this error probability is below 11\% for $\Delta < .784$, which (from eq.~(\ref{tilde_delta})) corresponds to $\tilde \Delta < .749$; this value should be compared to the value $\tilde\Delta=1$ for an oscillator in its vacuum state. Thus, secure squeezed-state key distribution is possible in principle using single-mode squeezed states, provided that the squeeze parameter $r$ defined by $\tilde\Delta=e^{-r}$ satisfies $r > -\log_e(.749) = .289$. When interpreted as suppression, relative to vacuum noise, of the quantum noise afflicting the squeezed observable, this amount of squeezing can be expressed as $10\cdot \log_{10} \left(\tilde\Delta^{-2}\right)= 2.51$ dB.

\begin{figure}
\begin{center}
\leavevmode
\epsfxsize=3in
\epsfbox{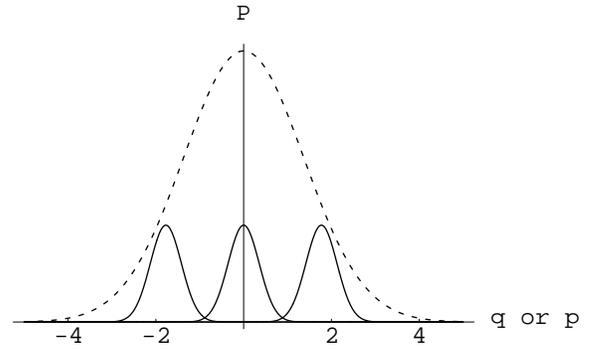}
\end{center}
\caption{Probability distributions for the squeezed quantum key distribution protocol, with squeeze factor $\tilde \Delta=1/2$. The dotted line is the probability distribution $P$ (a Gaussian with variance $(1/2\tilde \Delta^2)\cdot (1-\tilde \Delta^4)$) that Alice samples to determine the center of the squeezed signal that she sends. The solid lines are the probability distributions in position or momentum of the squeezed states (Gaussians with variance $\tilde \Delta^2/2$, shown with a different vertical scale than $P$) centered at $-\sqrt{\pi}$, 0, and $\sqrt{\pi}$. The intrinsic error probability due to imperfect squeezing (prior to binary error correction and privacy amplification) is  $1.2\%$.}
\label{fig:plot}
\end{figure}

The error rate is below $1\%$ for $\tilde \Delta < .483$ ($\Delta < .486$), and drops precipitously for more highly squeezed states, {\it e.g.}, to below $10^{-6}$ for $\tilde \Delta\sim \Delta < .256$. For example, if the noise in the channel is weak, Alice and Bob can use the Gaussian squeezed state protocol with $\tilde \Delta\sim 1/2$ (see Fig.~\ref{fig:plot}) to generate a shared bit via the $q$ or $p$ channel with an error rate ($\sim 1.2 \%$) comfortably below $11\%$; thus the protocol is secure if augmented with classical binary error correction and privacy amplification.

Of course, if the channel noise is significant, there will be a more stringent limit on the required squeezing. Many kinds of noise (for instance, absorption of photons in an optical fiber) will cause a degradation of the squeezing factor. If this is the only consequence of the noise, the squeezing exiting the channel should still satisfy $\Delta < .784$ for the protocol to be secure, as we discuss in more detail in \S\ref{sec:losses}. Otherwise, the errors due to imperfect squeezing must be added to errors from other causes to determine the overall error rate.

So far we have described the case where the $p$ states and the $q$ states are squeezed by equal amounts. The protocol works just as well in the case of unequal squeezing, if we adjust the error correction procedure accordingly. Consider carrying out the entanglement distillation using the code with general parameter $\alpha$ rather than $\alpha=1$. The error rates are unaffected if the squeezing in $q$ and $p$ is suitably rescaled, so that the width of the $q$ and $p$ states becomes
\begin{equation}
\Delta_q= \Delta\cdot \alpha~,\quad \Delta_p= \Delta /\alpha ~.
\end{equation}
In this modified protocol, Alice broadcasts the value of $q$ modulo $\sqrt{\pi}\cdot \alpha$ or the value of $p$ modulo $\sqrt{\pi}/\alpha$.  Bob subtracts the value broadcast by Alice from his own measurement outcome, and then adjusts the difference he obtains to the nearest multiple of $\sqrt{\pi}\cdot \alpha$ or $\sqrt{\pi}/\alpha$. The key bit is determined by whether the multiple of $\sqrt{\pi}\cdot \alpha$, or $\sqrt{\pi}/\alpha$, is even or odd. 

Thus, for example, the error rate sustained due to imperfect squeezing will have the same (acceptably small) value irrespective of whether Alice sends states with $\Delta_q=\Delta_p=1/2$, or $\Delta_q=1$ and $\Delta_p=1/4$; Alice can afford to send coherent states about half the time if she increases the squeezing of her other transmissions by a compensating amount.

Can we devise a secure quantum key distribution scheme in which Alice always sends coherent states? To obtain, as a reduction of an entanglement distillation protocol, a protocol in which coherent states ($\tilde \Delta=1$) are always transmitted, we must consider the case $\Delta^2=2$. But in that case, the initial state of Alice's and Bob's oscillators is a product state. Bob's value of $q$ or $p$ is completely uncorrelated with Alice's, and the protocol obviously won't work. This observation does not exclude secure quantum key distribution schemes using coherent states, but if they exist another method would be needed to prove the security of such schemes.

In general, the source that we obtain by measuring half of the entangled pair is biased. If $\Delta$ is not small compared to $\sqrt{\pi}$, then Alice is significantly more likely to generate a 0 than a 1 as her raw key bit. But as we have already discussed in \S\ref{subsec:bias}, after error correction and privacy amplification, the protocol is secure if $p_X$ and $p_Z$ are both less than $11\%$. This result follows because the squeezed state protocol is obtained as a reduction of an entanglement distillation protocol.

\section{Losses and other imperfections}
\label{sec:losses}

The ideal BB84 quantum key distribution protocol is provably secure. But in practical settings, the protocol cannot be implemented perfectly, and the imperfections can compromise its security. (See \cite{lutkenhaus} for a recent discussion.) For example, if the transmitted qubit is a photon polarization state carried by an optical fiber, losses in the fiber, detector inefficiencies, and dark counts in the detector all can impose serious limitations. In particular, if the photons travel a distance large compared to the attenuation length of the fiber, then detection events will be dominated by dark counts, leading to an unacceptably large error rate.

Furthermore, most present-day implementations of quantum cryptography use, not single photon pulses, but weak coherent pulses; usually the source ``emits'' the vacuum state, occasionally it emits a single photon, and with nonnegligible probability it emits two or more photons. Quantum key distribution with weak coherent pulses is vulnerable to a ``photon number splitting'' attack, in which the eavesdropper diverts extra photons, and acquires complete information about their polarization without producing any detectable disturbance. A weaker pulse is less susceptible to photon number splitting, but increases the risk that the detector will be swamped by dark counts. 

From a practical standpoint, quantum key distribution with squeezed states may not necessarily be better than BB84, but it is certainly different. Alice requires a source that produces a specified squeezed state on demand; fortunately, the amount of squeezing needed to ensure the security of the protocol is relatively modest. Bob uses homodyne detection to measure a specified quadrature amplitude; this measurement may be less sensitive to detector defects than the single-photon measurement required in BB84.

But, as in the BB84 protocol, losses due to the absorption of photons in the channel will enhance the error rate in squeezed-state quantum key distribution, and so will limit the distance over which secure key exchange is possible. We study this effect by modeling the loss as a damping channel described by the master equation
\begin{equation}
\dot \rho = \Gamma\left(a\rho a^\dagger - {1\over 2}a^\dagger a \rho - {1\over 2} \rho a^\dagger a\right) ~;
\label{eq:master}
\end{equation}
here $\rho$ is the density operator of the oscillator, $a$ is the annihilation operator, and $\Gamma$ is the decay rate. Eq.~(\ref{eq:master}) implies that
\begin{equation}
{d\over dt}\langle {a^\dagger}^k a^l\rangle_t = -{1\over 2}(k+l)\Gamma \langle {a^\dagger}^k a^l\rangle_t~,
\end{equation}
where
\begin{equation}
\langle {\cal O}\rangle_t= {\rm tr}~\left({\cal O}\rho(t)\right)
\end{equation}
denotes the expectation value of the operator ${\cal O}$ at time $t$. Integrating, we find
\begin{equation}
\langle{a^\dagger}^k a^l\rangle_{T}= e^{-{1\over 2}(k+l)\Gamma T}\langle {a^\dagger}^k a^l\rangle_{0}~,
\end{equation}
and so, by expanding in power series,
\begin{equation}
\langle :f(a^\dagger,a):\rangle_{T}= \langle :f(\xi a^\dagger,\xi a):\rangle_{0}~,\quad 
\xi=e^{-\Gamma T/2}
\label{normalevolve}
\end{equation}
where $f$ is an analytic function, and $:f:$ denotes normal ordering (that is, in $:f(a^\dagger,a):$, all $a^\dagger$'s are placed to the left of all $a$'s).

In particular, by normal ordering and applying eq.~(\ref{normalevolve}), we find 
\begin{equation}
\langle e^{i\beta q} \rangle_T = e^{-{1\over 4}(1-\xi^2)\beta^2}\langle e^{i\beta \xi q} \rangle_0~,
\label{gen_evolve}
\end{equation}
where $q=\left(a+a^\dagger\right)/\sqrt{2}$ is the position operator. A similar formula applies to the momentum operator or any other quadrature amplitude. Eq.~(\ref{gen_evolve}) shows that if the initial state at $t=0$ is Gaussian ($q$ is governed by a Gaussian probability distribution), then so is the final state at $t=T$ \cite{holevo}. The mean $\langle q\rangle$ and variance $\Delta q^2$ of the initial and final distributions are related by
\begin{eqnarray}
& &\langle q \rangle_T= \xi \langle q\rangle_0~,\nonumber\\
& &\left(\Delta q^2_T-{1\over 2} \right)= \xi^2 \left(\Delta q^2_0-{1\over 2}\right)~.
\label{change_prob}
\end{eqnarray}

Now let's revisit the analysis of \S\ref{sec:gaussian}, taking into account the effects of losses. We imagine that Alice prepares entangled pairs of oscillators in the state eq.~(\ref{psi_delta}), and sends one oscillator to Bob through the lossy channel; then they perform entanglement purification. This protocol reduces to one in which Alice prepares a squeezed state that is transmitted to Bob. In the squeezed-state protocol, Alice decides what squeezed state to send by sampling the probability distribution $P(q_A)$ given in eq.~(\ref{pqa}); if she chooses the value $q_A$, then she prepares and sends the state $|\psi(q_A)\rangle$ in eq.~(\ref{psi_qa}). When it enters the channel, this state is governed by the probability distribution
\begin{equation}
P(q_B|q_A)= {1\over \tilde \Delta \sqrt{\pi}}\exp\left(-(q_B-q_{B0})^2/\tilde\Delta^2\right)~,
\end{equation} 
and when Bob receives the state this distribution has, according to eq.~(\ref{change_prob}), evolved to 
\begin{equation}
P'(q_B|q_A)= {1\over \Delta' \sqrt{\pi}}\exp\left(-(q_B-q'_{B0})^2/\Delta'^2\right)~,
\end{equation}
where
\begin{eqnarray}
& &q'_{B0}= \xi q_{B0}\equiv \xi (1-\tilde \Delta^4)^{1/2}q_A~,\nonumber\\
& &\Delta'^2= \xi^2\tilde\Delta^2 + (1-\xi^2)~.
\end{eqnarray}
By integrating over $q_A$ in $P'(q_A,q_B)= P'(q_B|q_A)\cdot P(q_A)$, we can obtain the final marginal distribution for the difference $q_A-q_B$:
\begin{eqnarray}
\label{loss_difference}
& &P'(q_A-q_B;\xi)= {1\over \Delta_\xi \sqrt{\pi}}\exp\left(-(q_A-q_{B})^2/\Delta_\xi^2\right)~,\nonumber\\
& &\Delta_\xi^{-2}= {\tilde\Delta^2\over 1 +\xi^2 - 2\xi (1-\tilde\Delta^4)^{1/2} + (1-\xi^2)\tilde \Delta^2}~,
\end{eqnarray}
which generalizes eq.~(\ref{delta_tilde_delta}). We can express the damping factor $\xi$ as
\begin{equation}
\xi= e^{-\kappa d/2}~,
\end{equation}
where $d$ is the length of the channel and $\kappa^{-1}$ is its attenuation length (typically of the order of 10 km in an optical fiber).

The protocol is secure if the error rate in both bases is below $11\%$; as in \S\ref{sec:gaussian}, this condition is satisfied for $\Delta_\xi < .784$. Thus we can calculate, as a function of the initial squeezing parameter $\tilde \Delta$, the maximum distance $d_{\rm max}$ that the signal states can be transmitted without compromising the security of the protocol.

For $\tilde\Delta \ll 1$, we find
\begin{equation}
\kappa ~d_{\rm max} = (1.57)\cdot \tilde \Delta + O(\tilde \Delta^2)~.
\end{equation}
Thus, the more highly squeezed the input signal, the {\em less} we can tolerate the losses in the channel. This feature, which sounds surprising on first hearing, arises because the amount of squeezing is linked with the size of the range in $q_A$ that Alice samples. Errors are not unlikely if losses cause the value of $q_B$ to decay by an amount comparable to $\sqrt{\pi}/2$. In our protocol, if the squeezed states have a small width $\tilde \Delta$, then the typical states prepared by Alice are centered at a large value $q_A\sim \tilde\Delta^{-1}$; therefore, a small {\em fractional} decay can cause an error.

\begin{figure}
\begin{center}
\leavevmode
\epsfxsize=3in
\epsfbox{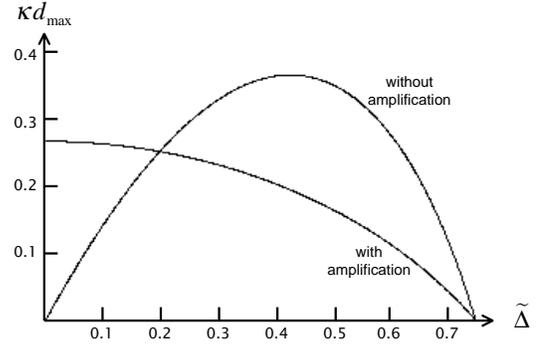}
\end{center}
\caption{The effect of channel losses on the security of quantum key distribution using squeezed states. The maximum length $\kappa d_{\rm max}$ of the channel (in units of the attenuation length) is plotted as a function of the width $\tilde \Delta$ of the squeezed state that enters the channel. For a longer channel, the error rate due to losses is too large and the proof of security breaks down. The curve labeled ``with amplification'' applies to the protocol in which the signal is amplified prior to detection in order to compensate for the losses; the curve labeled ``without amplification'' applies to the protocol in which the signal is not amplified.}
\label{fig:losses}
\end{figure}

On the other hand, even without losses, Alice needs to send states with $\tilde \Delta < .749$ to attain a low enough error rate, and as $\tilde \Delta$ approaches $.749$ from below, again only a small loss is required to push the error probability over 11\%. Thus there is an intermediate value of $\tilde \Delta$ that optimizes the value of $d_{\rm max}$, as shown in Fig.~\ref{fig:losses}. This optimal distance, 
\begin{equation}
\kappa ~d_{\rm max,opt}\approx .367~,
\end{equation}
is attained for $\tilde \Delta\sim .426$.

Our analysis so far applies if Alice and Bob have no prior knowledge about the properties of the channel. But if the loss $\xi^2=e^{-\kappa d}$ is known accurately, they might achieve a lower error rate if Bob compensates for the loss by multiplying his measurement outcome by $\xi^{-1}$ before proceeding with error correction and privacy amplification. This amplification of the signal by Bob is entirely classical, but to analyze the security in this case, we may consider an entanglement purification scenario in which Bob applies a quantum amplifier to the signal before measuring. Since the quantum amplifier (which amplifies all quadrature amplitudes, not just the one that Bob measures) is noisier, the protocol will be no less secure if Bob uses a classical amplifier rather than a quantum one.

So now we consider whether entanglement purification will succeed, where the channel acting on Bob's oscillator in each EPR pair consists of transmission through the lossy fiber followed by processing in Bob's amplifier. If the error rate is low enough, the key will be secure even if the amplifier, as well as the optical fiber, are under Eve's control.

Bob's linear amplifier can be modeled by a master equation like eq.~(\ref{eq:master}), but with $a$ and $a^\dagger$ interchanged, and where $\Gamma$ is now interpreted as a rate of gain. The solution is similar to eq.~(\ref{normalevolve}), except the normal ordering is replaced by {\em anti}-normal ordering (all $a$'s are placed to the {\em left} of all $a^\dagger$'s), and with $\xi^2$ replaced by the gain $\xi^{-2}=e^{\Gamma T}\ge 1$. We conclude that the amplifier transforms a Gaussian input state to a Gaussian output state, and that the mean $\langle q\rangle$ and variance $\Delta q^2$ of the Gaussian position distribution are modified according to
\begin{eqnarray}
&&\langle q\rangle \to \xi^{-1} \langle q\rangle~,\nonumber\\
&&\Delta q^2\to \xi^{-2}\Delta q^2 + {1\over 2}\left(\xi^{-2}-1\right)~.
\end{eqnarray}
Other quadrature amplitudes are transformed similarly.

Now suppose that a damping channel with loss $\xi^2$ is followed by an amplifier with gain $\xi^{-2}$. Then the mean of the position distribution is left unchanged, but the variance evolves as
\begin{eqnarray}
\Delta q^2&\to & \xi^{-2}\left(\xi^2\Delta q^2 + {1\over 2} \left(1-\xi^2\right)\right) + {1\over 2}\left(\xi^{-2}-1 \right)\nonumber\\
&=& \Delta q^2 +\left(\xi^{-2}-1\right)~.
\end{eqnarray}

For this channel, the probability distribution governing $q_A-q_B$ is again a Gaussian as in eq.~(\ref{loss_difference}), but now its width is determined by
\begin{equation}
\left(\Delta_\xi\right)_{\rm amp}^{-2}= {{1\over 2}\tilde\Delta^2\over 1 - (1-\tilde\Delta^4)^{1/2} + (\xi^{-2}-1)\tilde \Delta^2}~.
\end{equation}
Error rates in the $q$ and $p$ bases are below 11\%, and the protocol is provably secure, for $\left(\Delta_\xi\right)_{\rm amp}< .784$.

By solving $\left(\Delta_\xi\right)_{\rm amp}= .784$.
we can find the maximum distance $d$ (where $\xi^{-2}=e^{\kappa d}$) for which our proof of security holds; the result is plotted in Fig.~\ref{fig:losses}. When the squeezed input is narrow, $\tilde\Delta << 1$, the solution becomes 
\begin{equation}
\xi^{-2}\equiv\exp\left(\kappa ~d_{\rm max}\right)=1.307 + O(\tilde\Delta^2)~,
\end{equation}
or 
\begin{equation}
\kappa~ d_{\rm max}\approx .268~.
\end{equation}
Comparing the two curves in Fig.~\ref{fig:losses}, we see that the protocol with amplification remains secure out to longer distances than the protocol without amplification, {\em if} the input is highly squeezed. In that case, the error rate in the protocol without amplification is dominated by the decay of the signal, which can be corrected by the amplifier. But if the input is less highly squeezed, then the protocol without amplification remains secure to longer distances. In that case, the nonzero width of the signal state contributes significantly to the error rate; the amplifier noise broadens the state further. 

With more sophisticated protocols that incorporate some form of quantum error correction, 
continuous-variable quantum key distribution can be extended to longer distances. For example, if Alice and Bob share some noisy pairs of oscillators, they can purify the entanglement using protocols that require two-way classical communication \cite{plenio,zoller}. After pairs with improved fidelity are distilled, Alice, by measuring a quadrature amplitude in her laboratory, prepares a squeezed state in Bob's; the key bits can be extracted using the same error correction and privacy amplification schemes that we have already described. 

Our proof of security applies to the case where squeezed states are carried by a lossy channel (assuming a low enough error rate), because this scenario can be obtained as a reduction of a protocol in which Alice and Bob apply entanglement distillation to noisy entangled pairs of oscillators that they share. More generally, the proof applies to any imperfections that can be accurately modeled as a quantum operation that acts on the shared pairs before Alice and Bob measure them. As one example, suppose that when Alice prepares the squeezed state, it is not really the $q$ or $p$ squeezed state that the protocol calls for, but is instead slightly rotated in the quadrature plane. And suppose that when Bob performs his homodyne measurement, he does not really measure $q$ or $p$, but actually measures a slightly rotated quadrature amplitude. In the entanglement-distillation scenario, the imperfection of Alice's preparation can be modeled as a superoperator that acts on her oscillator before she makes a perfect quadrature measurement, and the misalignment of Bob's measurement can likewise be modeled by a superoperator acting on his oscillator before he makes a perfect quadrature measurement. Therefore, the squeezed state protocol with this type of imperfect preparation and measurement is secure, as long as the error rate is below 11\% in both bases. Of course, this error rate includes both errors caused by the channel and errors due to the imperfection of the preparation and measurement.

We also recall that in the protocols of \S\ref{sec:secure}, Alice's preparation and Bob's measurement were performed to $m$ bits of accuracy. In the entanglement distillation scenario, this finite resolution can likewise be well modeled by a quantum operation that shifts the oscillators by an amount of order $2^{-m}$ before Alice and Bob perform their measurements. Thus the proof applies, with the finite resolution included among the effects contributing to the permissible 11\% error rate.  The finite accuracy causes trouble only when Alice's and Bob's results lie a distance apart that is within about $2^{-m}$ of $\sqrt{\pi}/2$; thus, just a few bits of accuracy should be enough to make this additional source of error quite small.

\section{Conclusions}
\label{sec:conclude}
We have described a secure protocol for quantum key distribution based on the transmission of squeezed states of a harmonic oscillator. Conceptually, our protocol resembles the BB84 protocol, in which single qubit states are transmitted. The BB84 protocol is secure because monitoring the observable $Z$ causes a detectable disturbance in the observable $X$, and vice versa. The squeezed state protocol is secure because monitoring the observable $q$ causes a detectable disturbance in the observable $p$, and vice versa. Security is ensured even if the adversary uses the most general eavesdropping strategies allowed by the principles of quantum mechanics.

In secure versions of the BB84 scheme, Alice's source should emit single-photons that Bob detects. Since the preparation of single-photon states is difficult, and photon detectors are inefficient, at least in some settings the squeezed-state protocol may have practical advantages, perhaps including a higher rate of key production. Squeezing is also technically challenging, but the amount of squeezing required to ensure security is relatively modest.

The protocol we have described in detail uses each transmitted oscillator to carry one raw key bit. An obvious generalization is a protocol based on the code with stabilizer generators given in eq.~(\ref{n_and_alpha}), which encodes a $d$-dimensional protected Hilbert space in each oscillator. Then a secure key can be generated more efficiently, but more squeezing is required to achieve an acceptable error rate.

Our protocols, including their classical error correction and privacy amplification, are based on CSS codes: each of the stabilizer generators is either of the ``$q$''-type (the exponential of a linear combination of $n$ $q$'s) or of the ``$p$-type'' (the exponential of a linear combination of $n$ $p$'s). The particular CSS codes that we have described in detail belong to a restricted class: they are {\em concatenated} codes such that each oscillator encodes a single qubit, and then a block of those single-oscillator qubits are assembled to encode $k$ better protected qubits using a binary $[[n,k,d]]$ stabilizer code. There are more general CSS codes that embed $k$ protected qubits in the Hilbert space of $n$ oscillators but do not have this concatenated structure \cite{gott_kit_pres}; secure key distribution protocols can be based on these too. The quantum part of the protocol is still the same, but the error correction and privacy amplification make use of more sophisticated close packings of spheres in $n$ dimensions. 

We analyzed a version of the protocol in which Alice prepares Gaussian squeezed states governed by a Gaussian probability distribution. The states, and the probability distribution that Alice samples, need not be Gaussian for the protocol to be secure. However, for other types of states and probability distributions, the error rates might have to be smaller to ensure the security of the protocol.

Our proof of security applies to a protocol in which the squeezed states propagate through a lossy channel, over a distance comparable to the attentuation length of the channel. To extend continuous-variable quantum key distribution to much larger distances, quantum error correction or entanglement distillation should be invoked.

Strictly speaking, the security proof we have presented applies if Alice's state preparation (including the probability distribution that she samples) can be exactly realized by measuring half of an imperfectly entangled state of two oscillators. The protocol remains secure if Alice's source can be well approximated in this way. Our proof does not work if Alice occasionally sends two identically prepared oscillators when she means to send just one; the eavesdropper can steal the extra copy, and then the privacy amplification is not guaranteed to reduce the eavesdropper's information to an exponentially small amount. 

\acknowledgments

We thank Andrew Doherty, Steven van Enk, Jim Harrington, Jeff Kimble, and especially Hoi-Kwong Lo for useful discussions and comments. This work has been supported in part by the Department of Energy under Grant No. DE-FG03-92-ER40701, and by DARPA through the Quantum Information and
Computation (QUIC) project administered by the Army Research Office under Grant
No. DAAH04-96-1-0386. Some of this work was done at the Aspen Center for Physics.


\end{document}